\documentclass[11pt]{article}
\usepackage{amssymb}
\usepackage{amsmath}

\catcode`@=11
\renewcommand{\section}{\@startsection{section}{1}{0pt}{\medskipamount}
{\medskipamount}{\large\bf}}
\numberwithin{equation}{section}
\catcode`@=12

\def\a{\alpha}
\def\b{\beta}
\def\g{\gamma}
\def\de{\delta}
\def\eps{\epsilon}

\def\th{\theta}

\def\l{\lambda}
\def\m{\mu}
\def\n{\nu}
\def\r{\rho}
\def\s{\sigma}

\newcommand{\C}{\mathbb C}
\newcommand{\R}{\mathbb R}

\newcommand{\N}{\mathbb N}
\newcommand{\Hcal}{{\cal H}}

\def\e{\mbox{e}}
\def\i{\mbox{i}}
\def\N2{$N{=}2$}
\def\pa{\mbox{$\partial$}}
\def\diff{\mbox{d}}
\def\tr{{\rm tr}}
\def\sfrac#1#2{{\textstyle\frac{#1}{#2}}}
\def\rd#1{\buildrel{_{_{\hskip 0.01in}\rightarrow}}\over{#1}}
\def\ld#1{\buildrel{_{_{\hskip 0.01in}\leftarrow}}\over{#1}}

\newcommand{\adag}{a^{\dagger}}
\newcommand{\cdag}{c^{\dagger}}
\newcommand{\Tdag}{T^{\dagger}}
\newcommand{\Mdag}{M^{\dagger}}
\newcommand{\fh}{\hat{f}}
\newcommand{\gh}{\hat{g}}
\newcommand{\gb}{\overline{\gamma}}
\newcommand{\lb}{\overline{\lambda}}
\newcommand{\zb}{\overline{z}}
\newcommand{\wb}{\overline{w}}
\newcommand{\Pht}{\widetilde{\Phi}}
\newcommand{\Gt}{\widetilde{\Gamma}}
\newcommand{\Tt}{\widetilde{T}}
\newcommand{\Pt}{\widetilde{P}}
\newcommand{\Th}{\widehat{T}}
\newcommand{\Ph}{\widehat{P}}

\begin{document}
\begin{titlepage}
\setcounter{page}{0}
\begin{flushright}
hep-th/0106213\\
ITP--UH--04/01\\
\end{flushright}

\vskip 2.0cm

\begin{center}

{\Large\bf  Noncommutative Multi-Solitons in 2+1 Dimensions }

\vspace{14mm}

{\large Olaf Lechtenfeld\ and\ Alexander D. Popov~$^*$ }
\\[5mm]
{\em Institut f\"ur Theoretische Physik  \\
Universit\"at Hannover \\
Appelstra\ss{}e 2, 30167 Hannover, Germany }\\
{Email: lechtenf, popov@itp.uni-hannover.de}

\end{center}

\vspace{2cm}

\begin{abstract}
\noindent
The study of noncommutative solitons is greatly facilitated if the
field equations are integrable, i.e. result from a linear system.
For the example of a modified but integrable $U(n)$ sigma model
in $2{+}1$ dimensions we employ the dressing method to construct
explicit multi-soliton configurations on noncommutative~$\R^{2,1}$.
These solutions, abelian and nonabelian, feature exact time-dependence
for any value of the noncommutativity parameter~$\th$ and describe
various lumps of finite energy in relative motion. We discuss their
scattering properties and prove asymptotic factorization for large times.
\end{abstract}

\vfill

\textwidth 6.5truein
\hrule width 5.cm
\vskip.1in

{\small
\noindent ${}^*$
On leave from Bogoliubov Laboratory of Theoretical Physics, JINR,
Dubna, Russia}

\end{titlepage}

\section{Introduction}

\noindent
The past year has witnessed an explosion of activity in noncommutative
field theory. The original motivation derived from the discovery~\cite{seiberg}
that a certain corner of string moduli space is described by noncommutative
gauge theory. Since then, the field has been driven by the curiosity
to extend all kinds of (quantum) field theoretical structures to the
noncommutative realm~\cite{gopa} (for a review see~\cite{douglas}).

Before attempting to quantize noncommutative theories, it is certainly
warranted to achieve control over the moduli space of classical configurations.
Already {\it commutative\/} field theories in $1{+}1$, $2{+}1$, and $3{+}1$
dimensions display a variety of nonperturbative solutions (like solitons,
vortices, monopoles, instantons) which may be interpreted as D-branes in the
string context. Turning on a constant magnetic NS $B$-field background
generalizes those branes to solitonic solutions of the {\it noncommutatively\/}
deformed field theories~\cite{dasgupta,martinec} (for a lecture on the subject
see~\cite{komaba}).
Since spatial noncommutativity requires at least two dimensions, most
investigations have focused on scalar and gauge theories in $2{+}0$ and $2{+}1$
dimensions, where noncommutativity is tuned by a single parameter~$\th$.
One may roughly distinguish three types of results here.

Firstly, noncommutative {\it scalar\/} field theories were expanded around
their $\th{\to}\infty$ limit. There, the potential energy dominates, and
{\it limiting static solutions\/} are given by free projectors on subspaces
in a harmonic-oscillator Fock space. Corrections due to the kinetic energy can
be taken into account by a perturbation series in~$1/\th$
(see~\cite{gopa,jatkar,gopa2,rocek} and references therein).
Secondly, {\it explicit solutions\/} for finite~$\th$ were considered,
mainly for {\it abelian gauge\/} fields interacting with scalar matter
[2, 7 -- 23].  
Various properties of these configurations were investigated, including
their stability under quantum fluctuations
(see e.g.~\cite{aga,gross3,jackson} and references therein).
Thirdly, the {\it adiabatic motion\/} of solitons was studied by looking
at the geometry of the moduli space of the {\it static\/} multi-soliton
configurations (see e.g.~\cite{gopa2,rocek,araki} and references therein).

In a previous paper~\cite{LPS2} we proposed a noncommutative generalization
of an integrable sigma model with a Wess-Zumino-Witten-type term. This theory
describes the dynamics of $U(n)$-valued scalar fields in $2{+}1$ dimensions
\cite{ward}.
Its advantage lies in the fact that {\it multi-soliton\/} solutions to
its field equations can be written down explicitly.
This is possible because the equations of motion can be formulated as
the compatibility conditions of some linear equations.
Hence, this notion of integrability carries over to the noncommutative case,
and powerful solution-generating techniques can be applied here as well.

In the present paper we describe in detail how a new solution can be
constructed from an old one by a noncommutative generalization of the so-called
{\it dressing approach\/}~\cite{zakharov,zakh2,forgacs}
and write down a rather general class of explicit solutions,
{\it abelian\/} as well as {\it nonabelian\/}, for {\it any value of\/} $\th$.
These configurations are parametrized by arbitrary functions, and for concrete
choices we obtain (time-dependent) {\it multi-solitons\/}.
The latter feature finite-energy lumps in mutually {\it relative motion\/}.
We discuss their functional form, large-time asymptotics,
and scattering properties.

The paper is organized as follows.
We present the noncommutative extension of the modified sigma model,
its energy functional and field equations in the following Section.
Section~3 reviews some basics on the operator formalism of noncommutative
field theory. The dressing approach is outlined in Section~4, where we
follow our one-pole ansatz through to the general solution, which includes
the introduction of a `squeezing' transformation to moving-frame coordinates.
Abelian and nonabelian static solutions including their energy, BPS bound and
star-product form, are discussed in Section~5. We study one-soliton
configurations, i.e. solutions with a single velocity parameter, in Section~6,
derive their energies and give examples and limits. Section~7 finally
exemplifies the multi-soliton construction on the two-soliton case (with
relative motion), abelian as well as nonabelian. The asymptotic behavior at
high speed or large times is derived, which aids the evaluation of the energy.
In three appendices, we collect formulae about coherent and squeezed states
and present two different proofs of the large-time factorization of
multi-soliton configurations.

\section{Modified sigma model in 2+1 dimensions}

\noindent
{\bf Definitions and notation}.
As has been known for some time, nonlinear sigma models in $2{+}1$ dimensions
may be Lorentz-invariant or integrable but not both.
In this paper we choose the second property and investigate the
noncommutative extension of a modified $U(n)$ sigma model
(so as to be integrable) introduced by Ward~\cite{ward}.

Classical field theory on noncommutative spaces may be realized
in a star-product formulation or in an operator formalism.
The first approach is closer to the commutative field theory:
It is obtained by simply deforming the ordinary product of classical fields
(or their components) to the noncommutative star product
\begin{equation}
(f \star g)(x)\ =\ f(x)\,\exp\,\bigl\{ \frac{\i}{2}
{\ld{\partial}}_a \,\theta^{ab}\, {\rd{\partial}}_b \bigr\}\,g(x)\quad,
\end{equation}
with a constant antisymmetric tensor~$\th^{ab}$, where $a,b,\ldots=0,1,2$.
Specializing to $\R^{1,2}$, we shall use (real) coordinates $(x^a)=(t,x,y)$
in which the Minkowskian metric reads $(\eta_{ab})=\textrm{diag}(-1,+1,+1)$.
For later use we introduce two convenient coordinate combinations, namely
\begin{equation} \label{lightcone}
u\ :=\ \sfrac{1}{2}(t+y)\quad,\qquad
v\ :=\ \sfrac{1}{2}(t-y)\quad,\qquad
\pa_u\ =\ \pa_t+\pa_y\quad,\qquad
\pa_v\ =\ \pa_t-\pa_y
\end{equation}
and
\begin{equation}
z  \ :=\ x+\i y\quad,\qquad
\zb\ :=\ x-\i y\quad,\qquad
\pa_z    \ =\ \sfrac{1}{2}(\pa_x-\i \pa_y)\quad,\qquad
\pa_{\zb}\ =\ \sfrac{1}{2}(\pa_x+\i \pa_y)\quad.
\end{equation}
Since the time coordinate remains commutative, the only non-vanishing
component of the noncommutativity tensor is
\begin{equation}
\th^{xy}\ =\ -\th^{yx}\ =:\ \th\ >\ 0
\qquad\Longrightarrow\qquad
\th^{z\zb}\ =\ -\th^{\zb z}\ =\ -2\i\th \quad.
\end{equation}

\noindent
{\bf Action and energy}.
The noncommutative $U(n)$ sigma model may be obtained by a reduction 
of the Nair-Schiff sigma-model-type action~\cite{NaS,losev} 
from four to three dimensions,
\begin{align}
S\ &=\ -\frac12\int\!\diff{t}\,\diff{x}\,\diff{y}\;\eta^{ab}\;
\tr\,\Bigl(\pa_a \Phi^{-1} \star\, \pa_b \Phi \Bigr) \nonumber\\
&\quad\ - \frac13\int\!\diff{t}\,\diff{x}\,\diff{y} \int_0^1\!\diff{\r}\;
\widetilde{v}_{\l}\,\eps^{\l\m\n\s}\;\tr\,\Bigl(
\Pht^{-1}\star\,\pa_{\m}\Pht\,\star\,
\Pht^{-1}\star\,\pa_{\n}\Pht\,\star\,
\Pht^{-1}\star\,\pa_{\s}\Pht \Bigr) \quad,
\label{NaSaction}
\end{align}
where Greek indices include the extra coordinate~$\r$,
and $\eps^{\l\m\n\s}$ denotes the totally antisymmetric tensor in~$\R^4$.
The field~$\Phi(t,x,y)$ is group-valued, $\Phi^\dagger=\Phi^{-1}$,
with an extension $\Pht(t,x,y,\r)$ interpolating between
\begin{equation}
\Pht(t,x,y,0)\ =\ \textrm{const} \qquad\quad\textrm{and}\qquad\quad
\Pht(t,x,y,1)\ =\ \Phi(t,x,y) \quad,
\end{equation}
and `tr' implies the trace over the $U(n)$ group space.
Finally, $(\widetilde{v}_{\l})=(v_c,0)$
is a constant vector in (extended) space-time. For $(v_c)=(0,0,0)$
one obtains the standard (Lorentz-invariant) model.
Following Ward~\cite{ward}, we choose $(v_c)=(0,1,0)$ spacelike,
which yields a modified but integrable sigma model.
Although this explicitly breaks the Lorentz group of $SO(1,2)$ to the
$GL(1,\R)$ generated by the boost in $y$~direction, it leaves unmodified the
conserved energy functional
\begin{equation} \label{energy}
E\ =\ \frac{1}{2}\int\!\diff{x}\,\diff{y}\;\tr\,
\Bigl(\pa_t \Phi^{\dagger} \star \pa_t \Phi \,+\,
 \pa_x \Phi^{\dagger} \star \pa_x \Phi \,+\,
 \pa_y \Phi^{\dagger} \star \pa_y \Phi \Bigr) \quad.
\end{equation}

\noindent
{\bf Field equations}.
The noncommutative sigma-model equation of motion
following from~(\ref{NaSaction}) reads
\begin{equation}
(\eta^{ab}+v_c\,\eps^{cab})\,\pa_a (\Phi^{-1}\star\pa_b \Phi)\ =\ 0\quad,
\end{equation}
where $\eps^{abc}$ is the alternating tensor with $\eps^{012}{=}1$.
In our coordinates~(\ref{lightcone}) it is written more concisely as
\begin{equation} \label{yangtype}
\pa_x\,(\Phi^{-1}\star\pa_x\Phi)-\pa_v\,(\Phi^{-1}\star\pa_u\Phi)\ =\ 0 \quad.
\end{equation}
This Yang-type equation~\cite{Y} can be transformed
into a Leznov-type equation~\cite{L},
\begin{equation} \label{leznovtype}
\pa_x^2\phi -\pa_u\pa_v\phi +
\pa_v \phi \star \pa_x \phi - \pa_x \phi \star \pa_v \phi \ =\ 0 \quad ,
\end{equation}
for an algebra-valued field $\phi(t,x,y)\in u(n)$ defined via
\begin{equation} \label{Lax}
\pa_x\phi\ :=\ \Phi^{-1}\star\pa_u\Phi\ =:\ A
\qquad\textrm{and}\qquad
\pa_v\phi\ :=\ \Phi^{-1}\star\pa_x\Phi\ =:\ B \quad.
\end{equation}

The equation~(\ref{yangtype}) actually arises from the Bogomolnyi
equations for the $2{+}1$ dimensional Yang-Mills-Higgs system
$(A_\mu,\varphi)$,
\begin{equation}\label{bogo}
\sfrac{1}{2}\eps_{abc}\,F^{bc}\ =\
\pa_a \varphi + A_a \star\varphi - \varphi\star A_a \quad.
\end{equation}
Indeed, after choosing the gauge $A_v=0=A_x{+}\varphi$
and solving one equation by putting
\begin{equation}
A_u\ =\ \Phi^{-1} \star \pa_u \Phi \qquad\textrm{and}\qquad
A_x-\varphi\ =\ \Phi^{-1} \star \pa_x \Phi \quad,
\end{equation}
the Bogomolnyi equations (\ref{bogo}) get reduced to~(\ref{yangtype}).

\section{Operator formalism}

\noindent
{\bf Fock space}.
The nonlocality of the star product renders explicit computations cumbersome.
We therefore pass to the operator formalism,
which trades the star product for operator-valued coordinates
$\hat{x}^\mu$ satisfying $[\hat{x}^\mu,\hat{x}^\nu]=\i\theta^{\mu\nu}$.
The noncommutative coordinates for $\R^{1,2}$ are $(t,\hat{x},\hat{y})$
subject to
\begin{equation}
[t,\hat{x}]\ =\ [t,\hat{y}]\ =\ 0\quad, \qquad
[\hat{x},\hat{y}]\ =\ \i\theta
\qquad\Longrightarrow\qquad
[\hat{z},\hat{\zb}]\ =\ 2\theta \quad.
\end{equation}
The latter equation suggests the introduction of (properly normalized)
creation and annihilation operators,
\begin{equation} \label{adef}
a\ =\ \frac{1}{\sqrt{2\theta}}\,\hat{z} \qquad\textrm{and}\qquad
\adag\ =\ \frac{1}{\sqrt{2\theta}}\,\hat{\zb} \qquad \textrm{so that}\quad
[a,\adag]\ =\ 1 \quad.
\end{equation}
They act on a harmonic-oscillator Fock space $\Hcal$ with an orthonormal basis
$\{|n\rangle,\,n=0,1,2,\ldots\}$ such that
\begin{equation}
\adag a\,|n\rangle\ =:\ N\,|n\rangle\ =\ n\,|n\rangle \quad,\qquad
a\,|n\rangle\ =\ \sqrt{n}\,|n{-}1\rangle \quad, \qquad
\adag|n\rangle\ =\ \sqrt{n{+}1}\,|n{+}1\rangle \quad .
\end{equation}

\noindent
{\bf Moyal-Weyl map}.
Any function $f(t,z,\zb)$ can be related to an operator-valued
function $\fh(t)\equiv F(t,a,\adag)$ acting in $\Hcal$,
with the help of the Moyal-Weyl map (see e.g. \cite{alv,gross3})
\begin{align}
f(t,z,\zb)\quad \longrightarrow \quad F(t,a,\adag)\ &=\ -\int\!
\frac{\diff{p}\,\diff{\bar{p}}}{(2\pi)^2}\,\diff{z}\,\diff{\zb}\;
f(t,z,\zb)\;\e^{-\i[\bar{p}(\sqrt{2\theta}a-z)+p(\sqrt{2\theta}\adag-\zb)]}
\nonumber \\
&=\ \textrm{Weyl-ordered} \  f(t,\sqrt{2\th}\,a,\sqrt{2\th}\,\adag) \quad.
\end{align}
The inverse transformation recovers the c-number function,
\begin{align}
F(t,a,\adag)\ \longrightarrow\ f(t,z,\zb)\ &=\ 2\pi\theta \int\!
\frac{2\i\,\diff{p}\,\diff{\bar{p}}}{(2\pi)^2}\;\mbox{Tr}\Bigl\{F(t,a,\adag)\;
\e^{\i[\bar{p}(\sqrt{2\theta}a-z)+p(\sqrt{2\theta}\adag-\zb)]} \Bigr\}
\nonumber \\
&=\ F_\star \Bigl(
t,\frac{z}{\scriptstyle\sqrt{2\th}},\frac{\zb}{\scriptstyle\sqrt{2\th}}
\Bigr) \quad, \label{inverseMoyal}
\end{align}
where `Tr' signifies the trace over the Fock space~$\Hcal$,
and $F_\star$ is obtained from $F$ by replacing ordinary with star products.
Under the Moyal-Weyl map, we have
\begin{equation} \label{trace}
f\star g\ \longrightarrow\ \fh\,\gh \qquad\textrm{and}\qquad
\int\! \diff{x}\,\diff{y}\,f\ =\
2\pi \theta \,\mbox{Tr}\, \fh\ =\
2\pi \theta \sum_{n \geq 0} \langle n|\fh |n \rangle \quad.
\end{equation}
The operator formulation turns spatial derivatives into commutators,
\begin{equation}
\pa_x f \quad \longrightarrow \quad \frac{\i}{\theta}\, [\hat{y},\fh]
\ =:\ \hat{\pa}_x \fh
\qquad\textrm{and}\qquad
\pa_y f \quad \longrightarrow \quad -\frac{\i}{\theta}\, [\hat{x},\fh]
\ =:\ \hat{\pa}_y \fh \quad,
\end{equation}
so that
\begin{equation}
\pa_z f \quad \longrightarrow \quad \hat{\pa}_z \fh
\ =\ \frac{-1}{\sqrt{2\th}}\,[\adag,\fh]
\qquad\textrm{and}\qquad
\pa_{\zb} f \quad \longrightarrow \quad \hat{\pa}_{\zb} \fh
\ =\ \frac{1}{\sqrt{2\th}}\,[a,\fh] \quad.
\end{equation}
The basic examples for the relation between $f$ and $\hat{f}$ are
\begin{equation}
\begin{tabular}{cccccccc}
$f:$	   & $1$ & $z$ 	            & $\zb$	           & $z\zb{-}\th$ &
$z\zb$               & $z\zb{+}\th$  &  \\[4pt]
$\hat{f}:$ & $1$ & $\sqrt{2\th}\,a$ & $\sqrt{2\th}\,\adag$ & $2\th N$     &
$2\th(N{+}\sfrac12)$ & $2\th(N{+}1)$ & \quad. \\
\end{tabular}
\end{equation}
For more complicated functions it helps to remember that
\begin{equation}
z\star\zb\ =\ z\zb+\th \qquad\textrm{and}\qquad
\zb\star z\ =\ z\zb-\th \quad,\qquad
\end{equation}
because the composition law~(\ref{trace}) allows one to employ the
star product (on the functional side) instead of the Weyl ordering
(on the operator side).
For notational simplicity we will from now on omit the hats over the operators
except when confusion may arise.

\section{Dressing approach and explicit solutions}

\noindent
The payoff for considering an integrable model is the availability of
powerful techniques for constructing solutions to the equation of motion.
One of these tools is the so-called `dressing method', which was invented to
generate solutions for commutative integrable systems
\cite{zakharov,zakh2,forgacs}
and is easily extended to the noncommutative setup~\cite{LPS2}.
Let us briefly present this method (already in the noncommutative context)
before applying it to the modified sigma model.

\noindent
{\bf Linear system}.
We consider the two linear equations
\begin{equation}\label{linsys}
(\zeta \pa_x -\pa_u)\psi\ =\ A\,\psi \qquad\textrm{and}\qquad
(\zeta \pa_v -\pa_x)\psi\ =\ B\,\psi \quad,
\end{equation}
which can be obtained from the Lax pair for the (noncommutative) self-dual
Yang-Mills equations in $\R^{2,2}$ \cite{IP2,ivle}
by gauge-fixing and imposing the condition $\pa_3 \psi =0$.
Here, $\psi$ depends on $(t,x,y,\zeta)$ or, equivalently, on $(x,u,v,\zeta)$
and is an $n{\times}n$ matrix whose elements act as operators in the Fock
space $\Hcal$. The matrices $A$ and~$B$ are of the same type as $\psi$ but
do not depend on~$\zeta$.
The spectral parameter~$\zeta$ lies in the extended complex plane.
The matrix $\psi$ is subject to the following
reality condition~\cite{ward}:
\begin{equation}\label{real}
\psi(t,x,y,\zeta)\;[\psi(t,x,y,\bar{\zeta})]^{\dagger}\ =\ 1 \quad,
\end{equation}
where `$\dagger$' is hermitian conjugation. We also impose on $\psi$ the
standard asymptotic conditions \cite{ivle}
\begin{align}
\psi(t,x,y,\zeta\to\infty)\ &=\ 1\ +\ \zeta^{-1}\phi(t,x,y)\ +\ O(\zeta^{-2})
\quad, \label{asymp1} \\[4pt]
\psi(t,x,y,\zeta\to0)\ &=\ \Phi^{-1}(t,x,y)\ +\ O(\zeta)\quad. \label{asymp2}
\end{align}

The compatibility conditions for the linear system of differential equations
(\ref{linsys}) read
\begin{align}
\pa_x B -\pa_v A\ =\ 0 \quad ,\label{comp1} \\[4pt]
\pa_x A -\pa_u B -[A,B]\ =\ 0 \quad . \label{comp2}
\end{align}
We have already encountered `solutions' of these equations:
Expressing $A$ and $B$ in terms of $\phi$ like in~(\ref{Lax}),
\begin{equation} \label{Lax2}
A\ =\ \pa_x\phi\ =\ \Phi^{-1}\,\pa_u\Phi
\qquad\textrm{and}\qquad
B\ =\ \pa_v\phi\ =\ \Phi^{-1}\,\pa_x\Phi \quad,
\end{equation}
solves the first equation and turns the second into (the operator form of)
our Leznov-type equation~(\ref{leznovtype}),
\begin{equation} \label{leznovtype2}
\pa_x^2\phi -\pa_u\pa_v\phi -
[\pa_x \phi \,,\, \pa_v \phi] \ =\ 0 \quad.
\end{equation}
Alternatively, the ansatz employing $\Phi$ in~(\ref{Lax2})
fulfils the second equation and transforms the first one into (the operator
version of) our Yang-type equation~(\ref{yangtype}),
\begin{equation} \label{yangtype2}
\pa_x\,(\Phi^{-1}\,\pa_x\Phi)-\pa_v\,(\Phi^{-1}\,\pa_u\Phi)\ =\ 0 \quad.
\end{equation}
Inserting these two parametrizations of $A$ and $B$ into the linear system
(\ref{linsys}) we immediately verify (\ref{asymp1}) and (\ref{asymp2}),
confirming that the identical notation for $A$, $B$, $\phi$, and $\Phi$
in Section~2 and in the present one was justified.

\noindent
{\bf Dressing approach and ansatz}.
Having identified auxiliary linear first-order differential equations
pertaining to our second-order nonlinear equation, we set out to solve
the former. Note that the knowledge of~$\psi$ yields $\phi$ and $\Phi$
by way of (\ref{asymp1}) and (\ref{asymp2}), respectively,
and thus $A$ and $B$ via~(\ref{Lax2}) or directly from
\begin{align}
-\psi(t,x,y,\zeta)\;(\zeta\pa_x-\pa_u)[\psi(t,x,y,\bar{\zeta})]^{\dagger}\
=\ A(t,x,y) \quad , \label{A1} \\[4pt]
-\psi(t,x,y,\zeta)\;(\zeta\pa_v-\pa_x)[\psi(t,x,y,\bar{\zeta})]^{\dagger}\
=\ B(t,x,y) \quad \label{B1}.
\end{align}
The dressing method is a recursive procedure generating a new solution
from an old one. Let us have a solution $\psi_0(t,x,y,\zeta)$ of the
linear equations~(\ref{linsys}) for a given solution $(A_0,B_0)$
of the field equations (\ref{comp1}) and~(\ref{comp2}).
Then we can look for a new solution~$\psi$ in the form
\begin{equation} \label{generalansatz}
\psi(t,x,y,\zeta)\ =\ \chi(t,x,y,\zeta)\,\psi_0(t,x,y,\zeta)
\qquad\textrm{with}\qquad
\chi\ =\ 1\ +\ \sum_{\a=1}^s\sum_{k=1}^m\frac{R_{k\a}}{(\zeta-\mu_k)^\a}\quad,
\end{equation}
where the $\mu_k(t,x,y)$ are complex functions
and the $n{\times}n$ matrices $R_{k\a}(t,x,y)$ are independent of~$\zeta$.

In this paper we shall not discuss this approach in full generality
(for more details see e.g.~\cite{zakharov,zakh2,forgacs}).
We consider the vacuum seed solution
$A_0=B_0=0$, $\psi_0=1$ and a restricted ansatz containing only first-order
poles in~$\zeta$:
\begin{equation} \label{ansatzpsi}
\psi\ =\ \chi\,\psi_0\ =\ 1\ +\ \sum_{p=1}^m \frac{R_p}{\zeta-\mu_p} \quad,
\end{equation}
where the $\mu_p$ are complex constants with ${\rm Im}\mu_k<0$.
Moreover, following~\cite{forgacs}, we take the matrices~$R_k$ to be of the form
\begin{equation} \label{ansatzR}
R_k\ =\ \sum_{\ell=1}^m T_\ell\,\Gamma^{\ell k}\,\Tdag_k
\end{equation}
where the $T_k(t,x,y)$ are $n{\times}r$ matrices and $\Gamma^{\ell k}(t,x,y)$
are $r{\times}r$ matrices with some $r{\ge}1$.
In the abelian case ($n{=}1$) one may think of $T_k$ as a time-dependent
row vector
$\bigl(|z_k^1,t\rangle,|z_k^2,t\rangle,\ldots,|z_k^r,t\rangle\bigr)$
of (ket) states in~$\Hcal$.

We must make sure to satisfy the reality condition~(\ref{real}) as well as
our linear equations in the form (\ref{A1}) and~(\ref{B1}). In particular,
the poles at $\zeta=\bar{\mu}_k$ on the l.h.s. of these equations have to be
removable since the r.h.s. are independent of~$\zeta$. Inserting the ansatz
(\ref{ansatzpsi}) with (\ref{ansatzR}) and putting to zero the corresponding
residues, we learn from~(\ref{real}) that
\begin{equation} \label{Tequation}
\Bigl( 1-\sum_{p=1}^m \frac{R_p}{\mu_p-\bar{\mu}_k} \Bigr)\,T_k\ =\ 0
\end{equation}
while from (\ref{A1}) and (\ref{B1}) we obtain the differential equations
\begin{equation} \label{LRequation}
(A\ \textrm{or}\ B)\ =\
-\psi(\zeta)\,\overline{L}_k\,\psi(\bar{\zeta})^\dagger
\Big|_{\zeta\to\bar\mu_k}
\qquad\Longrightarrow\qquad
\Bigl( 1-\sum_{p=1}^m \frac{R_p}{\mu_p-\bar{\mu}_k} \Bigr)\,
\overline{L}_k\,R_k^\dagger\ =\ 0
\end{equation}
where
\begin{equation} \label{Ldef}
\overline{L}_k\ :=\ \bar{\mu}_k \pa_x - \pa_u
\qquad\textrm{or}\qquad
\overline{L}_k\ :=\ \bar{\mu}_k \pa_v - \pa_x \quad.
\end{equation}
The algebraic conditions~(\ref{Tequation}) imply that the $\Gamma^{\ell p}$
invert the matrices
\begin{equation}\label{Gamma}
\Gt_{pk}\ =\ \frac{1}{\mu_p{-}\bar{\mu}_k}\,\Tdag_p\,T_k
\quad, \qquad\textrm{i.e.}\qquad
\sum_{p=1}^m\,\Gamma^{\ell p}\,\Gt_{pk}\ =\
\delta_{\ k}^\ell \quad.
\end{equation}
Finally, by substituting (\ref{ansatzpsi}) and~(\ref{ansatzR}) into the
formulae (\ref{asymp2}) and~(\ref{asymp1}) we solve the equations of motion
(\ref{yangtype2}) and~(\ref{leznovtype2}) by
\begin{equation} \label{solform}
\Phi^{-1}\ =\ \Phi^{\dagger}\ =\
1\ -\ \sum_{k,\ell=1}^m \frac{1}{\mu_k}\,T_\ell\;\Gamma^{\ell k}\,T_k^\dagger
\qquad\qquad\textrm{and}\qquad\qquad
\phi\ =\ \sum_{k,\ell=1}^m T_\ell\;\Gamma^{\ell k}\,T_k^\dagger
\end{equation}
for every solution to (\ref{Tequation}) and~(\ref{LRequation}).

\noindent
{\bf Moving-frame coordinates}.
At this point it is a good idea to introduce the co-moving coordinates
\begin{align}
w_k\ &:=\ \nu_k\bigl[x+\bar{\mu}_k u +\bar{\mu}_k^{-1} v\bigr]\ =\
\nu_k\bigl[x + \sfrac{1}{2}(\bar{\mu}_k-\bar{\mu}_k^{-1})\,y +
\sfrac{1}{2}(\bar{\mu}_k+\bar{\mu}_k^{-1})\,t\bigr]
\nonumber \\[4pt] \label{wdef}
\wb_k\ &:=\ \bar\nu_k\bigl[x+\mu_k u +\mu_k^{-1} v\bigr]\ =\
\bar\nu_k\bigl[x + \sfrac{1}{2}(\mu_k-\mu_k^{-1})\,y +
\sfrac{1}{2}(\mu_k+\mu_k^{-1})\,t\bigr] \quad,
\end{align}
where the $\nu_k$ are functions of $\mu_k$ via
\begin{equation} \label{nudef}
\nu_k\ =\ \Bigl[\,\frac{4\i}{\mu_k-\bar\mu_k-\mu_k^{-1}+\bar\mu_k^{-1}}\,
\cdot\,\frac{\mu_k-\mu_k^{-1}-2\i}{\bar\mu_k-\bar\mu_k^{-1}+2\i}\,\Bigr]^{1/2}
\quad.
\end{equation}
In terms of the moving-frame coordinates the linear operators defined
in~(\ref{Ldef}) become
\begin{equation}
\overline{L}_k\ =\ \bar\nu_k\,(\bar{\mu}_k-\mu_k)\,\pa_{\wb_k}
\qquad\textrm{or}\qquad
\overline{L}_k\ =\ \bar\nu_k\,\mu_k^{-1} (\bar{\mu}_k-\mu_k)\,\pa_{\wb_k} \quad,
\end{equation}
respectively, so that we have only one equation~(\ref{LRequation})
for each pole. Since
\begin{equation}
[\hat{w}_k,\hat{\wb}_k]\ =\ \sfrac{\i}{2}\theta\,
\nu_k\bar\nu_k\,(\mu_k-\bar{\mu}_k-\mu_k^{-1}+\bar{\mu}_k^{-1})
\ =\ 2\,\theta
\end{equation}
we are led to the definition of co-moving creation and annihilation operators
\begin{equation}
c_k\ =\ \frac{1}{\sqrt{2\theta}}\,\hat{w}_k
\qquad\textrm{and}\qquad
\cdag_k\ =\ \frac{1}{\sqrt{2\theta}}\,\hat{\wb}_k
\qquad \textrm{so that}\quad
[c_k,\cdag_k]\ =\ 1 \quad.
\end{equation}
The static case ($w_k{\equiv}z$) is recovered for $\mu_k=-\i$.
Coordinate derivatives are represented in the standard fashion as
\cite{alv,gross3}
\begin{equation}\label{ccreat2}
\sqrt{2\theta}\,\pa_{w_k}
\quad\longrightarrow\quad -[\cdag_k, .\,] \quad, \qquad
\sqrt{2\theta}\,\pa_{\wb_k}
\quad\longrightarrow\quad [c_k, .\,] \quad.
\end{equation}
The number operator $N_k:=\cdag_k c_k$ is diagonal
in a co-moving oscillator basis $\bigl\{|n\rangle_k, n{=}0,1,2,\ldots\bigr\}$.

It is essential to relate the co-moving oscillators to the static one.
Expressing $w_k$ by $z$ via (\ref{wdef}) and~(\ref{nudef})
and using~(\ref{adef}) one gets
\begin{equation} \label{unitary}
c_k\ =\ (\cosh\tau_k)\,a - (\e^{\i\vartheta_k}\sinh\tau_k)\,\adag
- \b_k\,t\ =\ U_k(t)\,a\,U_k^\dagger(t) \quad,
\end{equation}
which is an inhomogeneous $SU(1,1)$ transformation mediated by the operator
\cite{perel}
\begin{equation} \label{Udef}
U_k(t)\ =\ \e^{\frac12\a_k\,a^{\dagger 2} -\frac12\bar\a_k\,a^2}\,
\e^{(\b_k\,\adag-\bar\b_k\,a)t} \quad.
\end{equation}
Here, we have introduced the complex quantities
\begin{equation}
\a_k\ =\ \e^{\i\vartheta_k}\,\tau_k
\qquad\textrm{and}\qquad
\b_k\ =\ -\sfrac12\nu_k\,(\bar\mu_k{+}\bar\mu_k^{-1})/\sqrt{2\th} \quad,
\end{equation}
with the relations
\begin{equation}
\xi_k\ :=\ \e^{\i\vartheta_k}\tanh\tau_k\ =\
\frac{\bar\mu_k{-}\bar\mu_k^{-1}{-}2\i}{\bar\mu_k{-}\bar\mu_k^{-1}{+}2\i}
\qquad\textrm{and}\qquad
\nu_k\ =\ \cosh\tau_k-\e^{\i\vartheta_k}\sinh\tau_k \quad,
\end{equation}
so that $\mu_k$ may be expressed in terms of~$\xi_k$ via
$\bar\mu_k=\i\frac{(1+\sqrt{\xi_k})^2}{1\,-\,\xi_k}$.
Clearly, we have $|n\rangle_k=U_k(t)|n\rangle$, and all co-moving
oscillators are unitarily equivalent to the static one.
The unitary transformation~(\ref{unitary}) is known
as `squeezing' (by $\a_k$) plus `shifting' (by $\b_k\,t$).

\noindent
{\bf Solutions}.
After this diversion, the remaining equations of motion~(\ref{LRequation})
take the form
\begin{equation} \label{eom}
\Bigl(1-\sum_{p=1}^m\frac{R_p}{\mu_p-\bar{\mu}_k}\Bigr)\,c_k\,T_k\ =\ 0\quad,
\end{equation}
which means that $c_k T_k$ lies in the kernel of the parenthetical expression.
Obviously, a sufficient condition for a solution is
\begin{equation} \label{holT}
c_k\,T_k\ =\ T_k\,Z_k
\end{equation}
with some $r{\times}r$ matrix $Z_k$.
We briefly elaborate on two important special cases.

Firstly, if $n{\ge}2$ and
\begin{equation}
Z_k\ =\ {\bf1}\,c_k \qquad\textrm{then}\qquad [c_k,T_k]\ =\ 0 \quad,
\end{equation}
which can be interpreted as a holomorphicity condition on~$T_k$.
Hence, any $n{\times}r$ matrix~$T_k$ whose entries are arbitrary functions
of~$c_k$ (but independent of~$c_k^\dag$) provides a solution~$\psi$ of the
linear system~(\ref{linsys}), after inserting it into (\ref{Gamma}),
(\ref{ansatzR}), and (\ref{ansatzpsi}). The complex moduli~$\mu_k$ turn
out to parametrize the velocities~$\vec{v}_k$ of individual asymptotic
lumps of energy. The shape and location of these lumps is encoded
in further moduli hiding in the functions $T_k(c_k)$.

Secondly, the abelian case ($n{=}1$) deserves some attention.
Here, the entries of the row vector
$T_k=\bigl(|z_k^1,t\rangle,|z_k^2,t\rangle,\ldots,|z_k^r,t\rangle\bigr)$
are {\it not\/} functions of~$c_k$. If we take
\begin{equation}
Z_k\ =\ \textrm{diag}\,(z_k^1,z_k^2,\ldots,z_k^r)
\quad\textrm{with}\quad z_k^i\in\C
\qquad\textrm{then}\qquad c_k\,|z_k^i,t\rangle\ =\ z_k^i\,|z_k^i,t\rangle\quad,
\end{equation}
which explains our labeling of the kets and qualifies them as coherent states
based on the co-moving ground state~$|0\rangle_k$,
\begin{equation}
|z_k^i,t\rangle\ =\ \e^{z_k^i\cdag_k-\zb_k^i c_k}|0\rangle_k\ =\
\e^{z_k^i\cdag_k-\zb_k^i c_k}\,U_k(t)\,|0\rangle\ =\
U_k(t)\,\e^{z_k^i\adag-\zb_k^i a}\,|0\rangle \quad.
\end{equation}
In total, the abelian solutions $|z_k^i,t\rangle$ are nothing but squeezed
states. The squeezing parameter is~$\a_k$
while the (complex) shift parameter is $z_k^i+\b_k t$.
In the star-product picture, we find $m$ groups of $r$ asymptotic lumps of
energy. The squeezing parametrizes the (common) deviation of the lumps of
type~$k$ from spherical shape, while the shifts yield the position of each
lump in the noncommutative plane. Clearly, all lumps in one group move
with equal constant velocity but are arbitrarily separated in the plane.
This discussion extends naturally to the nonabelian case.

We remark that our solutions have a four-dimensional perspective.
Recall that the linear system~(\ref{linsys}) can be produced from the linear
system of the self-dual Yang-Mills equations on noncommutative ${\R}^{2,2}$. 
Therefore, the dressing approach may be employed in this more general situation
to create BPS-type solutions of the functional form~(\ref{solform}) to the 
self-dual Yang-Mills equations in $2{+}2$ dimensions.  Hence, our soliton 
solutions of~(\ref{yangtype2}) in $2{+}1$ dimensions can be obtained from these
infinite-action BPS-type solutions by dimensional reduction.  In contrast, 
BPS-type solutions of the form~(\ref{solform}) do not exist on noncommutative
${\R}^{4,0}$ because the reality condition~(\ref{real}) for the matrix $\psi$
then involves $\zeta\to-\frac{1}{\bar\zeta}$ instead of $\zeta\to{\bar\zeta}$.
However, in four Euclidean dimensions a modified ADHM~approach allows one
to construct noncommutative instanton solutions the first examples of
which were given by Nekrasov and Schwarz~\cite{nekrasov}.

\section{Static solutions}

\noindent
{\bf Projectors}.
For comparison with earlier work, let us consider the static case,
$m=1$ and $\mu=-\i$.
Staying with our restricted ansatz of first-order poles only, the expression
(\ref{ansatzpsi}) then simplifies to
\begin{equation} \label{static}
\psi\ =\ 1\,+\,\frac{R}{\zeta+\i}\ =:\ 1\,-\,\frac{2\i}{\zeta+\i}\,P
\qquad\textrm{so that}\qquad \Phi^\dagger\ =\ 1\,-\,2\,P\ =\ P_\perp-P \quad.
\end{equation}
The reality condition~(\ref{real}) directly yields
\begin{equation}
P^\dagger\ =\ P \qquad\textrm{and}\qquad P^2\ =\ P \quad,
\end{equation}
qualifying $P$ as a hermitian projector.
Indeed, (\ref{real}) degenerates to $(1{-}P)P=0$,
while (\ref{LRequation}) simplifies to
\begin{equation} \label{eomstatic}
(1-P)\,\overline{L}\,P\ =\ 0 \qquad\Longrightarrow\qquad
F\ :=\ (1-P)\,a\,P\ =\ 0 \quad.
\end{equation}
We emphasize that this equation also characterizes the BPS subsector of
general noncommutative scalar field theories~\cite{gopa2,rocek}.

It is illuminating to specialize also our ansatz~(\ref{ansatzR})
for the residue, with~(\ref{Gamma}), to the static situation:
\begin{equation}
\Gamma\ =\ \frac{-2\i}{\Tdag T} \qquad\Longrightarrow\qquad
P\ \equiv\ \frac{R}{-2\i}\ =\ T\,\frac{\Gamma}{-2\i}\,\Tdag\ =\
T\,\frac{1}{\Tdag T}\,\Tdag \quad.
\end{equation}
Hence, the projector~$P$ is parametrized by an $n{\times}r$ matrix~$T$,
where, in the nonabelian case, $r{\le}n$ can be identified with
the rank of the projector in the $U(n)$ group space
(but not in~$\Hcal$).
The equation~(\ref{eom}) then simplifies to
\begin{equation} \label{eomstaticT}
(1-P)\,a\,T\ =\ 0 \quad,
\end{equation}
which means that $aT$ must lie in the kernel of~$1{-}P$.

\noindent
{\bf Energy and BPS bound}.
Clearly, within our ansatz static configurations are real, $\Phi^\dagger=\Phi$.
The energy~(\ref{energy}) of solutions to~(\ref{eomstatic}) reduces to
\begin{align} \label{energystatic}
E\ &=\ 4\pi \theta \,\mbox{Tr}\,\Bigl(
\hat{\pa}_z \Phi\,\hat{\pa}_{\zb} \Phi \Bigr)
\nonumber \\[4pt]
&=\ 8\pi\,\mbox{Tr}\,\Bigl( [\adag\,,P]\;[P\,,\,a] \Bigr)\
 =\ 8\pi\,\mbox{Tr}\,\Bigl( \adag Pa - P\,\adag a + 2\adag F \Bigr)
\nonumber \\[4pt]
&=\ 8\pi\,\mbox{Tr}\,\Bigl( \adag Pa - P\,\adag a \Bigr)\
\buildrel{?}\over{=}\ 8\pi\,\mbox{Tr}\,P \quad,
\end{align}
where the last equality holds only for
projectors of {\it finite\/} rank in~$\Hcal$.

The topological charge~$Q$ derives from a BPS argument.
The energy of a general static configuration
(within our ansatz but not necessarily a solution of $F{=}0$) obeys
\begin{align}
\sfrac{1}{8\pi}\,E\ &=\ \mbox{Tr}\,\Bigl([\adag,P]\,[P\,,a]\Bigr)\ =\
\mbox{Tr}\,\Bigl( P\,[\adag,P]\,[a\,,P]-P\,[a\,,P]\,[\adag,P] \Bigr) +
\mbox{Tr}\,\Bigl( F^\dagger F + F\,F^\dagger \Bigr)
\nonumber\\[6pt]  &\ge\
\mbox{Tr}\,\Bigl( P\,[\adag,P]\,[a\,,P]-P\,[a\,,P]\,[\adag,P] \Bigr)
\nonumber\\[6pt]  &\ =\
\sfrac{\i}{8}\,\theta\,\mbox{Tr}\,\Bigl(
\Phi\;\hat{\pa}_x\Phi\;\hat{\pa}_y\Phi\,-\,
\Phi\;\hat{\pa}_y\Phi\;\hat{\pa}_x\Phi \Bigr)
\ =:\ Q \quad.
\end{align}
Alternatively, using $G:=(1-P)\,\adag\,P$ in
\begin{equation}
\sfrac{1}{8\pi}\,E\ =\ \mbox{Tr}\,\Bigl([\adag,P]\,[P\,,a]\Bigr)\ =\ -
\mbox{Tr}\,\Bigl( P\,[\adag,P]\,[a\,,P]-P\,[a\,,P]\,[\adag,P] \Bigr) +
\mbox{Tr}\,\Bigl( G^\dagger G + G\,G^\dagger \Bigr)
\end{equation}
yields, along the same line, $E\ge-8\pi\,Q$.
In combination, we get
\begin{equation}
E\ \ge\ 8\pi\,|Q| \quad,
\qquad\textrm{with}\quad
E\ =\ 8\pi\,|Q| \quad\textrm{iff}\quad
F=0 \quad\textrm{or}\quad G=0 \quad.
\end{equation}
The $F{=}0$ configurations carry positive topological charge (solitons)
whereas the $G{=}0$ solutions come with negative values of~$Q$ (antisolitons).

It is instructive to imbed the BPS sector into the complete sigma-model
configuration space.
For static configurations, the sigma-model equation~(\ref{yangtype2}) becomes
\begin{equation} \label{yangtypeZ}
\pa_z\,(\Phi^{\dagger}\,\pa_{\zb}\Phi)+
\pa_{\zb}\,(\Phi^{\dagger}\,\pa_z\Phi)\ =\ 0
\qquad\buildrel{\Phi=1-2P}\over{\Longrightarrow}\qquad
[\,[a,[\adag,P]]\,,P\,]\ =\ 0 \quad.
\end{equation}
We emphasize that this is the equation of motion for the standard
noncommutative Euclidean two-dimensional grassmannian sigma model, as was 
clarified previously~\cite{LPS2}.  A short calculation shows that
solitons ($F{=}0$) as well as antisolitons ($G{=}0$) satisfy this equation.
To summarize, our choice of ansatz for~$\psi$ leads to the construction
of solitons. For describing antisolitons it suffices to switch from
$\mu_k$ to $\bar{\mu}_k$.

\noindent
{\bf Abelian static solutions}.
Abelian solutions ($n{=}1$) are now generated at ease.
Since $\mu=-\i$ implies $\a{=}0$ and $\b{=}0$,
the squeezing disappears and $r$ lumps are simply sitting at fixed positions
$z^i$, with $i{=}1,\ldots,r$. Hence, we have
\begin{equation}
T\ =\ \bigl(|z^1\rangle,|z^2\rangle,\ldots,|z^r\rangle\bigr)
\qquad\textrm{with}\quad
|z^i\rangle\ =\ \e^{z^i\adag-\zb^i a}\,|0\rangle \quad.
\end{equation}
This is obvious because $w\to z$ in the static case.
The coherent states~$|z^i\rangle$ are normalized to unity but are in general
not orthogonal.  One reads off the rank $r$ projector
\begin{equation}
P\ =\ \sum_{i,j=1}^r
|z^i\rangle\,\bigl(\langle z^\cdot|z^\cdot\rangle\bigr)_{ij}^{-1}\,\langle z^j|
 \ =\ {\cal U}\,\sum_{n=0}^{r-1} |n\rangle\langle n|\;{\cal U}^\dagger \quad,
\end{equation}
where the last equality involving a unitary operator~$\cal{U}$
was shown in~\cite{rocek}.
Clearly, the row vector
$T=\bigl(|0\rangle,|1\rangle,\ldots,|r{-}1\rangle\bigr)$
directly solves~(\ref{eomstaticT}),
and $P$ projects onto the space spanned by the $r$ lowest oscillator states.
Obviously, these solutions have $E=8\pi r$.

\noindent
{\bf Nonabelian static solutions}.
Another special case are nonabelian solutions with~$r{=}1$.
They are provided by column vectors~$T$ whose entries are functions
of~$a$ only.  If we choose polynomials of at maximal degree~$q$,
the energy of the configuration will be $E=8\pi q$, independently of
the gauge group.
To illustrate this fact we sketch the following $U(2)$ example:
\begin{equation} \label{u2ex}
T\ =\ \left(\begin{matrix} \l \\[10pt] a^q \end{matrix}\right)
\qquad\Longrightarrow\qquad
P\ =\  \left(\begin{matrix}
\frac{\l\lb}{\l\lb+N!/(N-q)!} & \frac{\l}{\l\lb+N!/(N-q)!}\,{\adag}^q \\[10pt]
a^q\,\frac{\lb}{\l\lb+N!/(N-q)!} & a^q\,\frac{1}{\l\lb+N!/(N-q)!}\,{\adag}^q
\end{matrix}\right) \quad,
\end{equation}
which is of infinite rank in~$\Hcal$. Still, the energy~(\ref{energystatic})
is readily computed with the result of~$8\pi q$.

\noindent
{\bf Inverse Moyal-Weyl map}.
Having constructed noncommutative solitons in operator language,
it is natural to wonder how these configurations look like in the
star-product formulation.
The back translation to functions~$\Phi_\star(t,x,y)$ solving the
noncommutative field equations~(\ref{yangtype}) is easily accomplished
by the inverse Moyal-Weyl map~(\ref{inverseMoyal}).
We take a brief look at $r{=}1$ solutions.

The abelian case has been considered many times in the literature
\cite{gopa,aga,harvey,komaba}. Here one has
\begin{equation} \label{abelex}
\Phi\ =\ 1\ -\ 2\,|0\rangle\langle0|
\qquad\longrightarrow\qquad
\Phi_\star\ =\ 1\ -\ 4\,\e^{-z\zb/\th} \quad,
\end{equation}
which lacks a nontrivial $\th{\to}0$ or $\th{\to}\infty$ limit.

For a nonabelian example we consider the static $U(2)$ solution~$\Phi=1-2P$
with $P$ given by~(\ref{u2ex}), for degree $q{=}1$.
With the methods outlined in Section~3, we find
\begin{equation} \label{u2ex2}
\Phi = 1 - 2\left(\begin{matrix}
\frac{\l\lb}{\l\lb+N} & \frac{\l}{\l\lb+N}\adag \\[10pt]
a\frac{\lb}{\l\lb+N} & a\frac{1}{\l\lb+N}\adag
\end{matrix}\right)
\qquad\longrightarrow\qquad
\Phi_\star = \left(\begin{matrix}
-\frac{\g\gb-z\zb+\th}{\g\gb+z\zb-\th} &
-2\g\frac{\g\gb+z\zb-2\th}{(\g\gb+z\zb-\th)^2}\zb \\[10pt]
-2 z\frac{\g\gb+z\zb-2\th}{(\g\gb+z\zb-\th)^2}\gb &
+\frac{\g\gb-z\zb-\th}{\g\gb+z\zb+\th}
\end{matrix}\right)
\end{equation}
with $\g=\sqrt{2\th}\l$.
One may check that it indeed fulfils~(\ref{yangtype}).

It is instructive to display the $\th\to0$ and $\th\to\infty$ limits
of~$\Phi_\star$, while keeping $\g$ fixed:
\begin{equation}
\Phi_\star(\th{\to}0)\ \to\ \left(\begin{matrix}
-\frac{\g\gb-z\zb}{\g\gb+z\zb} & \frac{-2\g\zb}{\g\gb+z\zb} \\[10pt]
 \frac{-2z\gb}{\g\gb+z\zb} & +\frac{\g\gb-z\zb}{\g\gb+z\zb}
\end{matrix}\right) \qquad\textrm{and}\qquad
\Phi_\star(\th{\to}\infty)\ \to\ \left(\begin{matrix}
+1 & 0 \\[10pt] 0 & -1 \end{matrix}\right) \quad.
\end{equation}
Other than the abelian solution, $\Phi_\star(z,\zb,\th)$ is a
rational function with a nontrivial commutative limit.

\section{One-soliton configurations}

\noindent

For a time-dependent configuration with $m{=}1$, our first-order pole
ansatz~(\ref{ansatzpsi}) simplifies to
\begin{equation} \label{movingpsi}
\psi\ =\ 1\,+\,\frac{R}{\zeta-\mu}\ =:\
1\,+\,\frac{\mu-\bar{\mu}}{\zeta-\mu}\,P
\qquad\textrm{with}\qquad
P\ =\ T\,\frac{1}{\Tdag T}\,\Tdag \quad,
\end{equation}
where $P$ and $T$ satisfy
\begin{equation} \label{movingP}
(1-P)\,c\,P\ =\ 0 \qquad\textrm{and}\qquad
c\,T = T\,Z \quad.
\end{equation}
Here,
\begin{equation} \label{movingc}
c\ =\ (\cosh\tau)\,a - (\e^{\i\vartheta}\sinh\tau)\,\adag - \b\,t
\ =\ U(t)\,a\,U^\dagger(t)
\end{equation}
is the creation operator in the moving frame.

Comparing the formulae of Section~4 with (\ref{movingpsi})--(\ref{movingc})
we see that the operators $c,\cdag$ and therefore the matrix~$T$ and
the projector~$P$ in~(\ref{movingP}) can be expressed in terms of the
corresponding static objects (see Section~5).
This is accomplished by means of the
inhomogeneous $SU(1,1)$ transformation~(\ref{movingc}),
which on the coordinates reads
\begin{equation} \label{boost}
\begin{pmatrix} w \\[4pt] \wb \end{pmatrix}\ =\
\begin{pmatrix} \cosh\tau & -\e^{\i\vartheta}\sinh\tau \\[4pt]
                -\e^{-\i\vartheta}\sinh\tau & \cosh\tau \end{pmatrix}
\begin{pmatrix} z \\[4pt] \zb \end{pmatrix}\ +\ \frac12
\begin{pmatrix} \nu\,(\bar\mu+\bar\mu^{-1}) \\[4pt]
                \bar\nu\,(\mu+\mu^{-1}) \end{pmatrix} \,t
\end{equation}
and is in fact a symplectic coordinate transformation,
preserving $\diff{z}\wedge\diff{\zb}$.
Even though the inhomogeneous part of~(\ref{boost}) is linear in time,
this does {\it not\/} mean that the solution
\begin{equation} \label{boostsol}
\Phi\ \equiv\ \Phi_{\vec v}\ =\ 1 - \bar\rho\,P
\qquad\textrm{with}\qquad \rho=1-\bar\mu/\mu
\end{equation}
is simply obtained by subjecting the corresponding static solution
$\Phi_{\vec 0}$ to such a `symplectic boost'.
To see this, consider the moving frame with the coordinates
$w,\wb$ and $t'{=}t$ and the related change of derivatives,
\begin{align}
\pa_x\ &=\ \nu\,\pa_w + \bar\nu\,\pa_{\wb} \quad, \nonumber \\[4pt]
\pa_t\ &=\ \sfrac12\nu\,(\bar{\mu}+\bar{\mu}^{-1})\,\pa_w +
           \sfrac12\bar\nu\,(\mu+\mu^{-1})\,\pa_{\wb} +
		   \pa_{t'} \quad, \nonumber \\[4pt]
\pa_y\ &=\ \sfrac12\nu\,(\bar{\mu}-\bar{\mu}^{-1})\,\pa_w +
           \sfrac12\bar\nu\,(\mu-\mu^{-1})\,\pa_{\wb} \quad. \label{ders}
\end{align}
In the moving frame our solution (\ref{boostsol}) will be static,
i.e.~$\pa_{t'}\Phi=0$. However, the field equation~(\ref{yangtype2})
is {\it not\/} invariant w.r.t. the coordinate transformation~(\ref{boost}).
Therefore, the solution~(\ref{boostsol}), which is static in the moving frame,
has a functional form different from that of the static solution~(\ref{static})
-- the coefficient of the projector is altered from $2$ to~$\bar\rho$.
Due to this non-invariance, the coordinate frame $(z,\zb,t)$ is distinguished
because in it the field equation takes the simplest form.

The constant velocity~$\vec v$ in the $xy$ plane is easily derived
by writing down the map~(\ref{boost}) in real coordinates
($w{=}{:}x'{+}\i y'$),~\footnote{
For geometric visualization,
the particular $SL(2,\R)$ transformation matrix appearing here may be
factorized to
$\begin{pmatrix} \cos\de & \sin\de \\ -\sin\de & \cos\de \end{pmatrix}
\begin{pmatrix} \cosh\tau&\sinh\tau\\ \sinh\tau&\cosh\tau \end{pmatrix}
\begin{pmatrix} \cos\de & -\sin\de \\ \sin\de & \cos\de \end{pmatrix}$,
where $2\de=-\frac{\pi}{2}-\vartheta$.
}
\begin{equation} \label{realboost}
\begin{pmatrix} x' \\[4pt] y' \end{pmatrix}\ =\
\begin{pmatrix}
\cosh\tau-\cos\vartheta\sinh\tau & -\sin\vartheta\sinh\tau \\[4pt]
-\sin\vartheta\sinh\tau & \cosh\tau+\cos\vartheta\sinh\tau
\end{pmatrix}
\biggl[
\begin{pmatrix} x \\[4pt] y \end{pmatrix}\ -\
\begin{pmatrix} v_x \\[4pt] v_y \end{pmatrix} \,t
\biggr] \quad,
\end{equation}
where we find
\begin{equation}\label{velo}
\vec{v}\ \equiv\ (v_x,v_y)\ =\
-\Bigl(\frac{\cos\varphi}{\cosh\eta}\ ,\ \frac{\sinh\eta}{\cosh\eta}\Bigr)
\qquad\textrm{for}\quad
\mu\ =:\ \e^{\eta-\i\varphi} \quad.
\end{equation}

The velocity dependence of the energy
\begin{equation} \label{energy2}
E\ =\ \pi \theta \,\mbox{Tr}\,\Bigl(
g^{ab}\,\pa_a\Phi_{\vec v}^\dagger\;\pa_b\Phi_{\vec v} \Bigr)
\qquad\textrm{with}\qquad
(g^{ab})\ =\ \textrm{diag}\,(+1,+1,+1)
\end{equation}
is now obtained straightforwardly by employing derivatives
w.r.t. $w$ and~$\wb$ via~(\ref{ders}), which yields
\begin{equation}
\begin{pmatrix} g^{w\wb} & g^{ww} \\[4pt] g^{\wb\wb} & g^{\wb w} \end{pmatrix}
\ =\ \frac12 \begin{pmatrix}
\nu\bar\nu(\sqrt{\mu\bar\mu}{+}\sqrt{\mu\bar\mu}^{-1})^2 &
\nu^2(\bar{\mu}{+}\bar{\mu}^{-1})^2 \\[4pt]
\bar\nu^2(\mu{+}\mu^{-1})^2 &
\nu\bar\nu(\sqrt{\mu\bar\mu}{+}\sqrt{\mu\bar\mu}^{-1})^2
\end{pmatrix} \quad.
\end{equation}
Inserting this bilinear form and (\ref{boostsol}) into the energy functional
(\ref{energy2}), we are left with~\cite{LPS2}
\begin{align}
E[\Phi_{\vec v}]\
&=\ \pi\theta\,\nu\bar\nu(\sqrt{\mu\bar\mu}{+}\sqrt{\mu\bar\mu}^{-1})^2\,
\mbox{Tr}\,\Bigl(\pa_w \Phi_{\vec v}^\dagger\,\pa_{\wb}\Phi_{\vec v} \Bigr)
\nonumber\\[4pt]
&=\ \pi\theta\,16 f(\vec{v})\,\mbox{Tr}\,
\Bigl( \pa_w P\,\pa_{\wb} P \Bigr) \nonumber\\[4pt]
&=\ 8\pi f(\vec{v})\,\mbox{Tr}\,\Bigl([\cdag,P]\,[P,c]\Bigr) \nonumber\\[4pt]
&=\ 8\pi f(\vec{v})\,q\ =\
f(\vec{v})\,E[\Phi_{\vec 0}] \quad, \label{boostE}
\end{align}
because the final trace is, by unitary equivalence, independent of the motion.
We have introduced the `velocity factor'
\begin{equation}
f(\vec{v})\ :=\ \sfrac{1}{16}\,\rho\bar\rho\,\nu\bar\nu\,
(\sqrt{\mu\bar\mu}+\sqrt{\mu\bar\mu}^{-1})^2\ =\
\frac{1}{\nu\bar\nu}\ =\
\cosh\eta\,\sin\varphi\ =\
\frac{\sqrt{1-\vec v^2}}{1-v_y^2} \quad.
\end{equation}
It is noteworthy that for motion in the $y$ direction ($\varphi{=}\frac\pi2$)
the `velocity factor'~$\cosh\eta$ agrees with the relativistic contraction factor
$1/\sqrt{1{-}\vec v^2}\ge1$, but it is smaller than one for motion with
$|v_x|>|v_y|\sqrt{1{-}\smash{v_y^2}}$, and the energy may be made arbitrarily
close to zero by a large `boost',
since $\vec v^2\to1$ is equivalent to $\varphi\to0$ or $\eta\to\pm\infty$.

\noindent
{\bf Examples}.
We can immediately write down moving versions of our solutions
(\ref{abelex}) and (\ref{u2ex2}) from the previous section.
Since $|0\rangle':=U(t)|0\rangle$ denotes the moving
ground state (i.e. squeezed and shifted), the $r{=}1$ abelian soliton reads
\begin{align}
\Phi\ =\ 1\ -\ \bar\rho\,U(t)\,|0\rangle\langle0|\,U(t)^\dagger
\qquad\Longrightarrow\qquad
\Phi_\star\ &=\ 1\ -\ 2\bar\rho\,\e^{-w\wb/\th} \nonumber\\
&=\ 1\ -\ 2\bar\rho\,
\e^{-[\vec r-\vec v t]^T \Lambda^T\!\Lambda\,[\vec r-\vec v t]\,/\th}
\quad,
\end{align}
where $\Lambda$ is the `squeezing matrix' appearing in~(\ref{realboost}).
In a similar vein, the $r{=}1$ nonabelian solution~(\ref{u2ex2})
gets generalized to
\begin{equation}
\Phi\ =\ 1\ -\ \bar\rho\,U(t) \left(\begin{matrix}
\frac{\l\lb}{\l\lb+\adag a} & \frac{\l}{\l\lb+\adag a}\adag \\[10pt]
a\frac{\lb}{\l\lb+\adag a} & a\frac{1}{\l\lb+\adag a}\adag
\end{matrix}\right) U(t)^\dagger\ =\
1\ -\ \bar\rho\,\left(\begin{matrix}
\frac{\l\lb}{\l\lb+\cdag c} & \frac{\l}{\l\lb+\cdag c}\cdag \\[10pt]
c\frac{\lb}{\l\lb+\cdag c} & c\frac{1}{\l\lb+\cdag c}\cdag
\end{matrix}\right) \quad,
\end{equation}
which leads to
\begin{equation}
\Phi_\star\ =\ 1\ -\ \bar\rho\,\left(\begin{matrix}
\frac{\g\gb}{\g\gb+w\wb-\th} &
\g\frac{\g\gb+w\wb-2\th}{(\g\gb+w\wb-\th)^2}\wb \\[10pt]
w\frac{\g\gb+w\wb-2\th}{(\g\gb+w\wb-\th)^2}\gb &
\frac{w\wb+\th}{\g\gb+w\wb+\th}
\end{matrix}\right) \quad,
\end{equation}
with the time dependence hiding in
\begin{equation}
w\ =\ w(t,x,y)\ =\
\bigl(\cosh\tau-\e^{\i\vartheta}\sinh\tau\bigr)(x-v_xt) + \i
\bigl(\cosh\tau+\e^{\i\vartheta}\sinh\tau\bigr)(y-v_yt) \quad,
\end{equation}
as expected.
We note that in the distinguished frame the large-time limit
(for $\vec v{\neq}\vec0$) is
\begin{equation}
\lim_{|t|\to\infty} \Phi_\star\ =\ 1\ -\ \bar\rho\,\Pi
\qquad\textrm{with}\qquad
\Pi\ =\ \begin{pmatrix} 0 & 0 \\ 0 & 1 \end{pmatrix} \quad.
\end{equation}

Although a one-soliton configuration with topological charge~$q{>}1$
may feature several separated lumps in its
energy density it does not deserve to be termed a `multi-soliton', because
those lumps do not display {\it relative\/} motion. True multi-solitons reduce
to a collection of solitons only in the asymptotic regime ($|t|\to\infty$)
or for large relative speed ($v_{\rm rel}\to\infty$);
they will be investigated next.

\section{Multi-soliton configurations}

\noindent
{\bf Two-soliton configurations}.
In Section~4 we gave an explicit recipe how to construct multi-soliton
configurations
\begin{equation}
\Phi^\dagger\ =\ 1\ -\ \sum_{k=1}^m R_k/\mu_k
\end{equation}
via (\ref{ansatzR}), (\ref{Gamma}), and (\ref{eom}).
It is known from the nonabelian commutative situation~\cite{ward}
that for polynomial functions $T_k(c_k)$ such solutions have finite energy
and describe $m$ different lumps of energy moving in the $xy$~plane with
distinct velocities $\vec v_k$ parametrized by
$\mu_k{=}\e^{\eta_k-\i\varphi_k}$, in other words: multi-solitons.
Since the explicit form of such configurations is more complicated than
that of one-soliton solutions, we illustrate the generic
features on the simplest examples, namely two-soliton configurations
with $r{=}1$. Thus, in the following, $m=2$ and $k=1,2$.

We begin by solving $\Gamma^{\ell p}$ in terms of $\Gt_{pk}$
via~(\ref{Gamma}), taking into account the noncommutativity:
\begin{align}
\Gamma^{11}\ &=\ +[ \Gt_{11} - \Gt_{12}\Gt_{22}^{-1}\Gt_{21} ]^{-1}
                  \quad,\qquad \phantom{\Gt_{12}\Gt_{22}^{-1}}
\Gamma^{12}\  =\ -[ \Gt_{11} - \Gt_{12}\Gt_{22}^{-1}\Gt_{21} ]^{-1}\,
                  \Gt_{12}\Gt_{22}^{-1} \quad, \nonumber \\
\Gamma^{21}\ &=\ -[ \Gt_{22} - \Gt_{21}\Gt_{11}^{-1}\Gt_{12} ]^{-1}\,
                  \Gt_{21}\Gt_{11}^{-1} \quad,\qquad
\Gamma^{22}\  =\ +[ \Gt_{22} - \Gt_{21}\Gt_{11}^{-1}\Gt_{12} ]^{-1} \quad,
\end{align}
which in terms of~$T_1$ and $T_2$ reads
\begin{align}
\Gamma^{11}\ &=\
+\Bigl[ \frac{1}{\mu_{11}} \,\Tdag_1 T_1 - \frac{\mu_{22}}{\mu_{12}\mu_{21}}\,
\Tdag_1 T_2 \frac{1}{\Tdag_2 T_2} \Tdag_2 T_1 \Bigr]^{-1} \quad,\nonumber\\
\Gamma^{12}\ &=\
-\Bigl[ \frac{1}{\mu_{11}} \,\Tdag_1 T_1 - \frac{\mu_{22}}{\mu_{12}\mu_{21}}\,
\Tdag_1 T_2 \frac{1}{\Tdag_2 T_2} \Tdag_2 T_1 \Bigr]^{-1}\,
\frac{\mu_{22}}{\mu_{12}} \,\Tdag_1 T_2 \frac{1}{\Tdag_2 T_2} \quad,\nonumber\\
\Gamma^{21}\ &=\
-\Bigl[ \frac{1}{\mu_{22}} \,\Tdag_2 T_2 - \frac{\mu_{11}}{\mu_{21}\mu_{12}}\,
\Tdag_2 T_1 \frac{1}{\Tdag_1 T_1} \Tdag_1 T_2 \Bigr]^{-1}\,
\frac{\mu_{11}}{\mu_{21}} \,\Tdag_2 T_1 \frac{1}{\Tdag_1 T_1} \quad,\nonumber\\
\Gamma^{22}\ &=\
+\Bigl[ \frac{1}{\mu_{22}} \,\Tdag_2 T_2 - \frac{\mu_{11}}{\mu_{21}\mu_{12}}\,
\Tdag_2 T_1 \frac{1}{\Tdag_1 T_1} \Tdag_1 T_2 \Bigr]^{-1} \quad,
\label{Gammasol}
\end{align}
with $\mu_{k\ell}:=\mu_k{-}\bar{\mu}_\ell$.
For the full solution, we then arrive at
\begin{align}
\Phi^\dagger\ =\ 1\ &-\
T_1 \Gamma^{11} \Tdag_1/\mu_1 - T_1 \Gamma^{12} \Tdag_2/\mu_2 -
T_2 \Gamma^{21} \Tdag_1/\mu_1 - T_2 \Gamma^{22} \Tdag_2/\mu_2 \nonumber\\[12pt]
=\ 1\
&-\ \frac{\mu_{11}}{\mu_1}\,T_1\,[\Tdag_1(1{-}\s P_2)T_1]^{-1}\,\Tdag_1\
 +\ \frac{\mu_{21}}{\mu_2}\s\,T_1\,[\Tdag_1(1{-}\s P_2)T_1]^{-1}\,\Tdag_1 P_2\
\nonumber\\[6pt]
&+\ \frac{\mu_{12}}{\mu_1}\s\,T_2\,[\Tdag_2(1{-}\s P_1)T_2]^{-1}\,\Tdag_2 P_1\
 -\ \frac{\mu_{22}}{\mu_2}\,T_2\,[\Tdag_2(1{-}\s P_1)T_2]^{-1}\,\Tdag_2
\label{twosolT} \\[12pt]
=\ 1\
&-\ \frac{1}{1{-}\s P_1 P_2}\,\Bigl\{
\frac{\mu_{11}}{\mu_1}\,P_1 - \frac{\mu_{21}}{\mu_2}\s\,P_1 P_2 \Bigr\}
 -\ \frac{1}{1{-}\s P_2 P_1}\,\Bigl\{
\frac{\mu_{22}}{\mu_2}\,P_2 - \frac{\mu_{12}}{\mu_1}\s\,P_2 P_1 \Bigr\}
\quad,
\label{twosolP}
\end{align}
where $\s:=\frac{\mu_{11}\mu_{22}}{\mu_{12}\mu_{21}}\in\R$,
the $T_k$ must be chosen to satisfy~(\ref{eom}),
and $P_k=T_k\frac{1}{\Tdag_k T_k}\Tdag_k$ as usual.

The abelian rank-one two-soliton is included here:
One simply writes
\begin{equation}
T_k\ =\ |z_k,t\rangle\ =\ \e^{z_k \cdag_k -\zb_k c_k}\,|0\rangle_k\ =\
\e^{z_k \cdag_k -\zb_k c_k}\,U_k(t)\,|0\rangle\ =\
U_k(t)\,\e^{z_k \adag -\zb_k a}\,|0\rangle \quad.
\end{equation}
Thus, noncommutative abelian fields look like commutative $U(\infty)$ fields.
Our kets are normalized to unity, and therefore
$P_k=|z_k,t\rangle\langle z_k,t|$.
In the bra-ket formalism our solution takes the form
\begin{align}
\Phi^\dagger\ =\ 1\ &-\
\frac{1}{1{-}\s|s|^2}\,\Bigl\{
\frac{\mu_{11}}{\mu_1}\,|z_1,t\rangle\langle z_1,t|\ -\
\frac{\mu_{21}}{\mu_2}\s s\,|z_1,t\rangle\langle z_2,t|
\nonumber\\[6pt] &\qquad\qquad\ -\
\frac{\mu_{12}}{\mu_1}\s \bar{s}\,|z_2,t\rangle\langle z_1,t|\ +\
\frac{\mu_{22}}{\mu_2}\,|z_2,t\rangle\langle z_2,t|\Bigr\} \quad,
\label{twosolabel}
\end{align}
where the overlap $\langle z_1,t|z_2,t\rangle$ was denoted by~$s$.

The simplest nonabelian example is a $U(2)$ configuration with
$T_1=(\begin{smallmatrix} 1 \\ a \end{smallmatrix})$ and
$T_2=(\begin{smallmatrix} 1 \\ c \end{smallmatrix})=UT_1 U^\dagger$,
i.e. taking $\mu_1=-\i$ and $\mu_2=:\mu$.
We refrain from writing down the lengthy explicit expression which
results from inserting this into (\ref{Gammasol}) and~(\ref{twosolT})
or from plugging the projectors $P_k=T_k\frac{1}{\Tdag_k T_k}\Tdag_k$
into~(\ref{twosolP}).

\noindent
{\bf High-speed asymptotics}.
The two-soliton configuration (\ref{twosolT}) does not factorize into
two one-soliton solutions, except in limiting situations,
such as large relative speed or large time.
Let us first consider the limit where the second lump
is moving with the velocity of light,
\begin{equation}
\varphi\ \to\ 0\quad\textrm{or}\quad\pi
\qquad\Longleftrightarrow\qquad
\textrm{Im}\,\mu_2\ \to\ 0
\qquad\Longleftrightarrow\qquad
|v_2|\ \to\ 1 \quad,
\end{equation}
while the velocity of the first lump is subluminal.
In this limit, we have $\s\to0$ and therefore
\begin{equation}
\Phi^\dagger\ \to\ 1\ -\
\frac{\mu_{11}}{\mu_1}\,P_1\ +\
\frac{\mu_{21}}{\mu_2}\s\,P_1 P_2\ +\
\frac{\mu_{12}}{\mu_1}\s\,P_2 P_1\ -\
\frac{\mu_{22}}{\mu_2}\,P_2 \quad.
\end{equation}
Define also $\rho_k:=\mu_{kk}/\mu_k=1{-}\bar{\mu}_k/\mu_k$.
Clearly, Im $\mu_2\to0$ implies that
\begin{equation}
\Phi^\dagger\ \to\ 1\ -\ \rho_1\,P_1 \quad,
\end{equation}
except for $\vec v_2\to(0,\pm1)$ meaning $\mu_2\to0$ or $\infty$, where
\begin{equation}
\Phi^\dagger\ \to\ (1 - \rho_1 P_1) (1 - \rho_2 P_2) \qquad\textrm{or}\qquad
\Phi^\dagger\ \to\ (1 - \rho_2 P_2) (1 - \rho_1 P_1) \quad,
\end{equation}
respectively.

\noindent
{\bf Large-time asymptotics}.
Now we investigate the behavior of~(\ref{twosolT}) for large (positive
and negative) time. For convenience, we take the first lump to be at rest
while the second one moves, i.e.
\begin{equation} \label{specialmu}
\mu_1=-\i \quad\textrm{and}\quad \mu_2=:\mu
\qquad\Longrightarrow\qquad
\rho_1=2 \quad,\quad \rho_2=:\rho \quad,\quad
\s=1-\frac{(\bar{\mu}-\i)(\mu+\i)}{(\mu-\i)(\bar{\mu}+\i)} \quad.
\end{equation}
For simplicity, we consider $r{=}1$, i.e. $T_k$ is a column vector
and $P_k$ of (matrix) rank one.
After some tedious algebra (see Appendix~B) one learns that $\Phi^\dagger$
factorizes in the large-time limit:
\begin{equation} \label{larget}
\Phi^{\dagger}\ \to\ \bigl(1-2\,\Pt_1\bigr)\,\bigl(1-\rho\,\Pi_2\bigr) \quad,
\end{equation}
where $\Pi_2:=\lim_{|t|\to\infty}P_2$ is a coordinate-independent hermitian
projector, and
\begin{equation}
\Pt_1\ =\ \bigl[\a\,\Pi_2+(1{-}\Pi_2)\bigr]\,
T_1\,\frac{1}{\Tdag_1(1{-}\s\Pi_2)\,T_1}\,\Tdag_1\,
\bigl[\bar{\a}\,\Pi_2+(1{-}\Pi_2)\bigr]
\qquad\textrm{with}\quad \a=\frac{\bar{\mu}-\i}{\mu-\i}
\end{equation}
is a new projector describing a one-soliton configuration.

We have explicitly checked this behavior for the simple $U(2)$ example
mentioned above and found indeed
\begin{equation}
\Phi^\dagger(x,y,t\to\pm\infty)\ =\
\bigl(1-2\,\Pt_1\bigr)\,\bigl(1-\rho\,\Pi_2\bigr)\ +\ O(t^{-1})
\qquad\textrm{with}\qquad
\Pt_1\ =\ \Tt_1\frac{1}{\Tt_1^\dagger\Tt_1}\Tt_1^\dagger \quad,
\end{equation}
where
\begin{equation}
\Tt_1\ =\ \begin{pmatrix} 1 \\ \a\,a \end{pmatrix}
\qquad\textrm{and}\qquad
\Pi_2\ =\ \begin{pmatrix} 0 & 0 \\ 0 & 1 \end{pmatrix} \quad.
\end{equation}

Let us now consider the situation within the frame moving with the second lump.
In this frame the second lump is static. By the same arguments as above,
the solution $\Phi^\dagger$ behaves in the large-time limit as
\begin{equation} \label{larget2}
\Phi^{\dagger}\ \to\ \bigl(1-2\,\Pi_1\bigr)\,\bigl(1-\rho\,\Pt_2\bigr) \quad,
\end{equation}
where $\Pi_1$ is a coordinate-independent hermitian projector,
and the projector $\Pt_2$ describes the second lump in the moving frame.
It is obvious how these statements generalize to $m$-soliton configurations.
If we sit in the frame moving with the $\ell$th lump, then the
large-time limit of~$\Phi^\dagger$ is a product of $(1-\rho_\ell\Pt_\ell)$ and
constant unitary matrices,
\begin{equation} \label{mlarget}
\Phi^\dagger\ \to\
(1-\rho_1\Pi_1)\,(1-\rho_2\Pi_2)\ldots(1-\rho_{\ell-1}\Pi_{\ell-1})\,
(1-\rho_\ell\Pt_\ell)\,(1-\rho_{\ell+1}\Pi_{\ell+1})\ldots(1-\rho_m\Pi_m) \;.
\end{equation}
The proof is outlined in Appendix~C.

A lesson from~(\ref{mlarget}) concerns the scattering properties of
our multi-soliton solutions. Since the limits $t\to-\infty$ and $t\to+\infty$
in~(\ref{mlarget}) are identical, every lump escapes completely unharmed
from the encounter with the other ones. The absence of any change in velocity,
shape, or displacement evidences the no-scattering feature of these lumps,
be they commutative or not.
Summarizing, what survives
at $|t|\to\infty$ is at most a one-soliton configuration
with modified parameters and multiplied with constant unitary matrices.

The asymptotic study just performed covers the $U(1)$ case as well,
with some additional qualifications.
At rank one and $m{=}2$, the static lump is parametrized by
\begin{align}
|z_1,0\rangle\ &\equiv\ |z_1\rangle\ =\ \e^{z_1 \adag-\zb_1 a}\,|0\rangle
\quad,\qquad\textrm{while} \nonumber\\[6pt]
|z_2,t\rangle\ &=\ \e^{z_2 \cdag-\zb_2 c}\,U(t)\,|0\rangle\ =\
U(t)\,\e^{z_2 \adag-\zb_2 a}\,|0\rangle
\end{align}
representing the moving lump remains time-dependent as $|t|\to\infty$.
Yet, the analysis is much simpler than for the nonabelian case.
Since~\cite{perel}
\begin{equation}
\langle z_1,0|z_2,t\rangle\ =\
\e^{-\frac12|\b_2|^2\,t^2}\,(1{-}\sfrac12|\xi_2|^2) \quad,
\end{equation}
the overlap $s=\langle z_1,0|z_2,t\rangle$ goes to zero exponentially fast
and the solitons become well separated.
Therefore, (\ref{twosolabel}) tends to
\begin{equation}
\Phi^\dagger\ \to\ 1\ -\
2\,|z_1\rangle\,\langle z_1|\ -\ \rho\,|z_2,t\rangle\,\langle z_2,t| \quad,
\end{equation}
which decomposes additively (rather than multiplicatively) into two
independent solitons.

\noindent
{\bf Energies.}
It is technically difficult to compute the energy
of true multi-solution configurations such as~(\ref{twosolT}).
However, because energy is a conserved quantity, we may evaluate it in
the limit of~$|t|\to\infty$.
{}From the result~(\ref{mlarget}) we seem to infer that generically
only the $\ell$th soliton contributes to the total energy.
Yet, this is incorrect, as can be seen already in the
commutative case: Even though the moving lumps have disappeared from sight
at $|t|=\infty$, their energies have not, because the integral of the energy
density extends to spatial infinity, and so it is not legitimate to
perform the large-time limit before the integration.
What~(\ref{mlarget}) shows, however, is the large-time vanishing of the
{\it overlap\/} between the lumps in relative motion, so that asymptotically
the lumps pertaining to different values of~$k$ are well separated.
The total energy, being a local functional, then approaches a sum of $m$
contributions, each of which stems from a group of $r$ lumps, in isolation
from the other groups.

To compute the $\ell$th contribution, we must ignore the influence of the
other groups of lumps. To this end we take the large-time limit and treat
the $\ell$th group as a moving one-soliton.
{}From (\ref{boostE}) we know that its energy is
\begin{equation} \label{energyell}
E_\ell(\vec v_\ell)\ =\ f(\vec v_\ell)\,E_\ell(\vec 0)\ =\
8\pi\, f(\vec v_\ell)\,q_\ell \quad.
\end{equation}
By adding all contributions,
it follows that the total energy of the $m$-soliton solution at any time is
\begin{equation}
E\ =\ 8\pi\,\sum_{k=1}^m
q_k\,\cosh\eta_k\,\sin\varphi_k  \quad.
\end{equation}

\section{Conclusions}

\noindent
In this paper we have demonstrated that the power of integrability can
be extended to noncommutative field theories without problems.
For the particular case of a $2{+}1$ dimensional integrable sigma model
which captures the BPS sector of the $2{+}1$ dimensional Yang-Mills-Higgs
system, the `dressing method' was applied to generate a wide class
of multi-soliton solutions to the noncommutative field equations.
We were able to transcend the limitations of standard noncommutative
scalar and gauge theories, by providing a general construction scheme for
and several examples of explicit analytical multi-soliton configurations,
with {\it full time dependence\/}, for {\it finite\/} $\th$ values, and
for any $U(n)$ {\it gauge group\/}. Moreover, two proofs of large-time
asymptotic factorization into a product of single solitons were presented,
and the energies and topological charges have been computed.

The model considered here does not stand alone, but is motivated
by string theory. As was explained in two previous papers~\cite{LPS1,LPS2},
$n$ coincident D2-branes in a $2{+}2$ dimensional space-time give rise to
open $N{=}2$ string dynamics~[42 -- 46] which (on tree level~\footnote{
for a one-loop analysis see~\cite{CLN,CS} }) is completely described
by our integrable $U(n)$ sigma model living on the brane. In this context, the
multi-solitons are to be regarded as D0-branes moving inside the D2-brane(s).
Switching on a constant NS~$B$-field and taking the Seiberg-Witten decoupling
limit deforms the sigma model noncommutatively, admitting also regular
abelian solitons.

Since the massless mode of the open $N{=}2$ string in a space-time-filling
brane parametrizes self-dual Yang-Mills in $2{+}2$ dimensions~\cite{OV},
it is not surprising that the sigma-model field equations also derive from the
self-duality equations by dimensional reduction~\cite{ward}.
Moreover, it has been shown that most (if not all) integrable equations in
three and less dimensions can be obtained from the self-dual Yang-Mills
equations (or their hierarchy) by suitable reductions
(see e.g. \mbox{[49 -- 53]} and references therein).
This implies, in particular, that open $N{=}2$ strings can provide a
consistent quantization of integrable models in $1{\le}D{\le}3$ dimensions.
Adding a constant $B$-field background yields {\it noncommutative\/} self-dual
Yang-Mills on the world volumes of coincident D3-branes and, hence,
our {\it noncommutative\/} modified sigma model on the world volumes of
coincident D2-branes~\cite{LPS1}.
It is certainly of interest to study also other reductions of the
noncommutative self-dual Yang-Mills equations to noncommutative versions of
KdV, nonlinear Schr\"odinger or other integrable equations, as have been
considered e.g. in~\cite{taka,legare,dimakis,paniak}.

Another feature of these models is a very rich symmetry structure.
All symmetries of our equations can be derived
(after noncommutative deformation) from symmetries of the self-dual Yang-Mills
equations~\cite{PP,po} or from their stringy generalizations~\cite{ivle,pole3}.
For this one has to extend the relevant infinitesimal symmetries and integrable
hierarchies to the noncommutative case. In fact, such
symmetries can be realized as derivatives w.r.t. moduli parameters
entering the solutions described in Sections 5 to~7 or, in a more general case,
as vector fields on the moduli space.

In this paper we considered the simultaneous and relative motion of several
noncommutative lumps of energy. Like in the commutative limit,
it turned out that {\it no scattering\/} occurs for the noncommutative
solitons within the ansatz considered here. This means that the different
lumps of energy move linearly in the spatial plane with transient shape
deformations at mutual encounters but asymptotically emerge without
time delay or changing profile or velocity.
However, starting from the ansatz~(\ref{generalansatz})
containing {\it second-order\/} poles in~$\zeta$, one can actually construct explicit
solutions featuring {\it nontrivial scattering\/} of noncommutative solitons.
Such configurations will generalize earlier results obtained for commutative
solitons~[58 -- 61] of the modified sigma model in $2{+}1$ dimensions.

It would also be very interesting to construct the Seiberg-Witten map for
our soliton configurations along the lines presented in~\cite{hashi}.
This should allow one to compare their representation in commutative
variables with their profile in the star-product formulation.

Finally, an exciting relation between the algebraic structures of
noncommutative soliton physics and string field theory has surfaced recently
\cite{kost,rast,gross4}. It is tempting to speculate that the `sliver states'
will play a role in the analysis of the open $N{=}2$ string field theory
\cite{berk,pole3}. We intend to investigate such questions in the near future.

\bigskip
\noindent
{\large{\bf Acknowledgements}}

\smallskip
\noindent
O.L. acknowledges a stimulating discussion with P. van Nieuwenhuizen.
This work is partially supported by DFG under grant Le~838/7-1.

\bigskip

\setcounter{section}{0}
\renewcommand{\thesection}{\Alph{section}}
\section{Coherent and squeezed states}

\noindent
We collect some definitions and useful relations pertaining to
coherent and squeezed states~\cite{perel}.

Coherent states for the Heisenberg algebra $[a,\adag]=1$ are generated
by the unitary shift operator
\begin{equation}
D(\b)\ :=\ \e^{\b \adag - \bar\b a}\ =\
\e^{-\frac12\bar\b\b}\,\e^{\b \adag}\,\e^{-\bar\b a} \quad.
\end{equation}
It acts on the Heisenberg algebra as a shift,
\begin{equation}
D\,\begin{pmatrix} a \\ \adag \end{pmatrix}\,D^\dagger\ =\
\begin{pmatrix} a \\ \adag \end{pmatrix} -
\begin{pmatrix} \b \\ \bar\b \end{pmatrix} \quad,
\end{equation}
and operates on the Fock space via
\begin{align}
D(\b)\,|0\rangle\ &=\ \e^{-\frac12\bar\b\b}\,\e^{\b \adag}\,|0\rangle \quad,\\
\langle0|\,D^\dagger(\b_1)\,D(\b_2)\,|0\rangle \ &=\
\e^{-\frac12\bar\b_1\b_1}\,\e^{-\frac12\bar\b_2\b_2}\,\e^{\bar\b_1\b_2} \quad.
\end{align}

A representation of $su(1,1)$ can be constructed in terms of
bilinears in the Heisenberg algebra,
\begin{equation}
K_+:=\sfrac12 {\adag}^2 \quad,\qquad
K_-:=\sfrac12 a^2 \quad,\qquad
K_0:=\sfrac14 (a\adag + \adag a) \quad,
\end{equation}
which fulfill
\begin{equation}
[K_0,K_\pm] = \pm K_\pm \qquad\textrm{and}\qquad
[K_+,K_-] = -2 K_0 \quad.
\end{equation}
The unitary squeezing operator $S$ may be defined as
\begin{equation}
S(\a)\ :=\ \e^{\a K_+ - \bar\a K_-} \ =\
\e^{\xi K_+}\,\e^{-\ln(1-\bar\xi\xi) K_0}\,\e^{-\bar\xi K_-}\
=:\ S(\xi) \quad,
\end{equation}
where
\begin{equation}
\a\ =:\ \e^{\i\vartheta}\,\tau \qquad\textrm{and}\qquad
\xi\ =\ \e^{\i\vartheta}\,\tanh\tau \quad.
\end{equation}
$S$ induces a representation on the Heisenberg algebra via
\begin{equation}
S\,\begin{pmatrix} a \\ \adag \end{pmatrix}\,S^\dagger\ =\
g\,\begin{pmatrix} a \\ \adag \end{pmatrix}
\qquad\textrm{with}\qquad
g\ =\ \begin{pmatrix}
\cosh\tau & -\e^{\i\vartheta}\sinh\tau \\[4pt]
-\e^{-\i\vartheta}\sinh\tau & \cosh\tau \end{pmatrix} \quad.
\end{equation}
The action of $S$ on the Fock space is characterized by
\begin{align} \nonumber
S(\xi)\,|0\rangle\ &=\ (1-\bar\xi\xi)^{1/4}\,e^{\xi K_+}\,|0\rangle \quad,\\
\langle0|\,S^\dagger(\xi_1)\,S(\xi_2)\,|0\rangle \ &=\
(1-\bar\xi_1\xi_1)^{1/4}\,(1-\bar\xi_2\xi_2)^{1/4}\,(1-\bar\xi_1\xi_2)^{-1/2}
\quad.
\end{align}

The combined action of shifting and squeezing leads to squeezed states
of the form
\begin{equation}
|\xi,\b\rangle\ :=\ S(\xi)\,D(\b)\,|0\rangle\ =\
\e^{-\frac12\bar\b\b}\,(1-\bar\xi\xi)^{1/4}\,\e^{\frac12\xi{\adag}^2+\b\adag}\,
|0\rangle \quad,
\end{equation}
which are normalized to unity due to the unitarity of $S$ and~$D$.
The corresponding transformation of the Heisenberg generators reads
\begin{equation}
\begin{pmatrix} c \\ \cdag \end{pmatrix}\ :=\ S\,D\,
\begin{pmatrix} a \\ \adag \end{pmatrix}\,D^\dagger\,S^\dagger\ =\
\begin{pmatrix}
\cosh\tau & -\e^{\i\vartheta}\sinh\tau \\
-\e^{-\i\vartheta}\sinh\tau & \cosh\tau \end{pmatrix}
\begin{pmatrix} a \\ \adag \end{pmatrix}\ -\
\begin{pmatrix} \b \\ \bar\b \end{pmatrix} \quad,
\end{equation}
where the order of action was relevant.
It is obvious that the unitary transformations defined in~(\ref{unitary})
are generated by $U_k(t)=S(\xi_k)D(\b_k t)$ with parameters $\{\xi_k,\b_k\}$
determined by~$\mu_k$.

\section{Asymptotic factorization for the two-soliton configuration}

\noindent
We set out to proof the result~(\ref{larget}).
With $\mu_1{=}{-}\i$ and $\mu_2{=}\mu{\neq}{-}\i$, we first consider the
distinguished $(z,\zb,t)$~coordinate frame.
Let the entries of~$T_2$ be polynomials of (the same) degree~$q_2$ in~$c_2$.
At large times, (\ref{unitary}) implies that $c_2\to-\b_2\,t$,
and the $t^{q_2}$ term in each polynomial will dominate,
\begin{equation} \label{T2limit}
T_2\ \to\ t^{q_2}
\left(\begin{smallmatrix}
\g_1 \\ \g_2 \\ \vdots \\ \g_n
\end{smallmatrix}\right)
\ =:\ t^{q_2}\,\vec\g \quad,
\end{equation}
where $\vec\g$ is a fixed vector in group space.
Due to the homogeneity of~(\ref{twosolT}),
the proportionality factor~$t^{q_2}$ can be dropped
and $\vec\g$ be normalized to unity, $\vec{\g}^{\;\dagger}\vec{\g}=1$.
We may then substitute
\begin{equation} \label{P2limit}
\Tdag_2\,T_2\ \to\ 1 \qquad\textrm{and}\qquad
P_2\ \to\ T_2\,\Tdag_2\ \to\ \vec{\g}\ \vec{\g}^{\;\dagger}
\ =:\ \Pi \ =\ \Pi^\dagger \quad,
\end{equation}
the projector onto the $\vec\g$ direction.
Note that the entries of the constant matrix $\Pi$ are commuting.
Furthermore, we introduce $p:=\vec{\g}^{\;\dagger}P_1\,\vec{\g}$
which is a group scalar but noncommuting.
Performing these substitutions in~(\ref{twosolT}),
the two-soliton configuration becomes
\begin{align}
\Phi^{\dagger}\ \to\ 1\
&-\ \frac{\mu_{11}}{\mu_1}\,
  T_1\,\frac{1}{\Tdag_1(1{-}\s\Pi)T_1}\,\Tdag_1\
 +\ \frac{\mu_{21}}{\mu_2}\s\,
  T_1\,\frac{1}{\Tdag_1(1{-}\s\Pi)T_1}\,\Tdag_1\,\Pi
\nonumber\\[6pt]
&+\ \frac{\mu_{12}}{\mu_1}\s\,\frac{1}{1{-}\s p}\,\Pi\,P_1\
 -\ \frac{\mu_{22}}{\mu_2}\,\frac{1}{1{-}\s p}\,\Pi \quad.
\end{align}
In order to turn the terms in the first line to projectors,
it is useful to introduce a constant matrix
\begin{equation}
M\ =\ \a\,\Pi + (1{-}\Pi) \qquad\textrm{with}\quad
\a=\frac{\bar{\mu}-\i}{\mu-\i} \qquad\textrm{so that}\qquad
\Mdag M = 1 - \s\Pi
\end{equation}
and rewrite
\begin{equation}
M\,T_1\ =:\ \Tt \qquad\textrm{so that}\qquad
T_1\,\frac{1}{\Tdag_1 \Mdag M\,T_1}\,\Tdag_1\ =\ \frac1M \Pt \frac1{\Mdag}
\qquad\textrm{with}\quad
\Pt := \Tt\,\frac{1}{\Tt^\dagger\Tt}\,\Tt^\dagger \quad.
\end{equation}
We thus arrive at
\begin{equation}
\Phi^{\dagger}\ \to\ 1\
-\ \frac{\mu_{11}}{\mu_1}\,\frac1M \Pt \frac1{\Mdag}\
+\ \frac{\mu_{21}}{\mu_2}\s\,\frac1M \Pt \frac1{\bar{\a}}\,\Pi\
+\ \frac{\mu_{12}}{\mu_1}\s\,\frac{1}{1{-}\s p}\,\Pi\,P_1\
 -\ \frac{\mu_{22}}{\mu_2}\,\frac{1}{1{-}\s p}\,\Pi \quad,
\end{equation}
after making use of
\begin{equation}
f(M)\,\Pi\ =\ \Pi\,f(M)\ =\ f(\a)\,\Pi \qquad\textrm{but}\qquad
f(M)\,(1{-}\Pi)\ =\ (1{-}\Pi)\,f(M)\ =\ 1{-}\Pi \quad.
\end{equation}
Further analysis is facilitated by decomposing $\Phi^{\dagger}$
into parts orthogonal and parallel to~$\vec{\g}$.
We insert our values~(\ref{specialmu}) for the coefficients and find
\begin{align} \label{decomp}
(1{-}\Pi)\,\Phi^\dagger\,(1{-}\Pi)\ &\to\
(1{-}\Pi)\,\Bigl(1-2\,\Pt\Bigr)\,(1{-}\Pi)\,
\\[12pt]
(1{-}\Pi)\,\Phi^\dagger\,\vec{\g}\ &\to\
(1{-}\Pi)\,\Bigl(-2\,\Pt\,\frac{1}{\bar{\a}} +
\frac{2\i(\mu{-}\bar{\mu})}{\mu(\bar{\mu}{+}\i)}\,\Pt\,
                \frac{1}{\bar{\a}}\Bigr)\,\vec{\g}
\nonumber\\[12pt]
\vec{\g}^{\;\dagger}\,\Phi^\dagger\,(1{-}\Pi)\ &\to\
\vec{\g}^{\;\dagger}\,\Bigl(-2\,\frac{1}{\a}\,\Pt +
\frac{2(\mu{-}\bar{\mu})}{\mu{-}\i}\,\frac{1}{1{-}\s p}\,P_1\Bigr)\,(1{-}\Pi)
\nonumber\\[12pt]
\vec{\g}^{\;\dagger}\,\Phi^\dagger\,\vec{\g}\ &\to\
\vec{\g}^{\;\dagger}\,\Bigl(1-2\,\frac{1}{\a}\,\Pt\,\frac{1}{\bar{\a}} +
\frac{2\i(\mu{-}\bar{\mu})}{\mu(\bar{\mu}{+}\i)}\,
                \frac{1}{\a}\,\Pt\,\frac{1}{\bar{\a}} +
\frac{2(\mu{-}\bar{\mu})}{\mu{-}\i}\,\frac{p}{1{-}\s p} -
\frac{\mu{-}\bar{\mu}}{\mu}\,\frac{1}{1{-}\s p}\Bigr)\,\vec{\g} \ .
\nonumber
\end{align}
With the help of the identities
\begin{align}
\vec{\g}^{\;\dagger}\,\Pt\,\vec{\g}\ &=\
\a\,\vec{\g}^{\;\dagger}\,\frac{1}{1-\s P_1\Pi}\,P_1\,\vec{\g}\,\bar{\a}\ =\
\a\,\bar{\a}\,\frac{p}{1-\s p} \qquad\qquad\textrm{and}
\nonumber\\[12pt]
\vec{\g}^{\;\dagger}\,\Pt\,(1{-}\Pi)\ &=\
\a\,\vec{\g}^{\;\dagger}\,\frac{1}{1-\s P_1\Pi}\,P_1\,(1{-}\Pi)\ =\
\a\,\frac{1}{1-\s p}\;\vec{\g}^{\;\dagger}\,P_1\,(1{-}\Pi)
\end{align}
plus some algebra,
the pieces of $\Phi^{\dagger}$ in (\ref{decomp}) simplify to
\begin{align}
(1{-}\Pi)\,\Phi^\dagger\,(1{-}\Pi)\ &\to\
(1{-}\Pi)\,\bigl(1-2\,\Pt\bigr)\,(1{-}\Pi)\,
\nonumber\\[8pt]
(1{-}\Pi)\,\Phi^\dagger\,\vec{\g}\ &\to\
(1{-}\Pi)\,\bigl(-2\,\Pt + 2\,\rho\,\Pt \bigr)\,\vec{\g}
\nonumber\\[8pt]
\vec{\g}^{\;\dagger}\,\Phi^\dagger\,(1{-}\Pi)\ &\to\
\vec{\g}^{\;\dagger}\,\bigl(-2\,\Pt \bigr)\,(1{-}\Pi)
\nonumber\\[8pt]
\vec{\g}^{\;\dagger}\,\Phi^\dagger\,\vec{\g}\ &\to\
\vec{\g}^{\;\dagger}\,\bigl(1-\rho-2\,\Pt+2\,\rho\,\Pt \bigr)\,\vec{\g} \quad,
\end{align}
which proves the factorization~\footnote{
We have omitted writing `soliton indices' for $\Pt$ and $\Pi$, to avoid
cluttering our formulae.}
\begin{equation} 
\Phi^{\dagger}\ \to\ \bigl(1-2\,\Pt\bigr)\,\bigl(1-\rho\,\Pi\bigr) \quad.
\end{equation}
At $|t|\to\infty$ we are thus left with a static one-soliton configuration
with modified parameters ($T_1\to MT_1$) and multiplied with a constant matrix
($1{-}\rho\Pi$).
If we move with the {\it second\/} lump the proof is the same, because then 
$a\to\b_2 t$ and $P_1$ (in place of~$P_2$) goes to a constant projector.

\section{Asymptotic form of m-soliton configurations}

\noindent
The dressing method may be used to show the large-time factorization
of multi-soliton solutions into a product of one-soliton solutions
directly on the level of the auxiliary function~$\psi(t,x,y,\zeta)$.

Starting from the one-soliton ansatz,
\begin{equation}
\psi_1\ =\ 1\ +\ \frac{\mu_1{-}\bar\mu_1}{\zeta{-}\mu_1}\,\Ph_1 
\end{equation}
with a hermitian projector~$\Ph_1$,
we may try to obtain a two-soliton solution by `dressing' the latter with
another factor of the same kind ($\mu_1\neq\mu_2$),
\begin{equation} \label{psi2}
\psi_2\ =\ \Bigl( 1+\frac{\mu_1{-}\bar\mu_1}{\zeta{-}\mu_1}\,\Ph_1 \Bigr)\,
           \Bigl( 1+\frac{\mu_2{-}\bar\mu_2}{\zeta{-}\mu_2}\,\Ph_2 \Bigr)\quad.
\end{equation}
The reality condition~(\ref{real}) is identically fulfilled as soon as
$\Ph_1$ and $\Ph_2$ are hermitian projectors,
\begin{equation}
\Ph_1\ =\ \Th_1\,\frac{1}{\Th_1^\dagger\Th_1}\,\Th_1^\dagger
\qquad\textrm{and}\qquad
\Ph_2\ =\ \Th_2\,\frac{1}{\Th_2^\dagger\Th_2}\,\Th_2^\dagger \quad.
\end{equation}
The removability of the pole for $\zeta\to\bar\mu_2$ in (\ref{A1}) and
(\ref{B1}) is guaranteed if
\begin{equation} \label{hatequation}
(1-\Ph_2)\,c_2\,\Ph_2\ =\ 0
\qquad\Longrightarrow\qquad
c_2\,\Th_2\ =\ \Th_2\,Z_2 \quad,
\end{equation}
qualifying the second factor on the r.h.s. of (\ref{psi2}) 
as a standard one-soliton solution, as obtained from solving~(\ref{holT}).
We therefore redenote $\Ph_2\to\Pt_2$ and $\Th_2\to\Tt_2$.~\footnote{
It will be seen shortly that these objects coincide with the ones defined
in Appendix~B.}
The vanishing of the $\zeta\to\bar\mu_1$ residue, on the other hand, leads to
\begin{equation} \label{Ph1equation}
(1-\Ph_1)\,\Bigl(1+\frac{\mu_2{-}\bar\mu_2}{\bar\mu_1{-}\mu_2}\,\Pt_2\Bigr)\,
c_1\,\Bigl(1+\frac{\bar\mu_2{-}\mu_2}{\bar\mu_1{-}\bar\mu_2}\,\Pt_2\Bigr)\,\Ph_1
\ =\ 0 \quad,
\end{equation}
which does not yield a `holomorphicity condition' for $\Th_1$ but rather 
demonstrates that $\Ph_1$ is {\it not\/} a standard one-soliton projector.

Let us move with the first lump.
In the $|t|\to\infty$ limit, then, the arguments used in (\ref{T2limit})
and~(\ref{P2limit}) apply, meaning that $\Pt_2\to\Pi_2$.
Since $\Pi_2$ is a coordinate-independent projector, 
the $c_1$ in~(\ref{Ph1equation}) can be moved next to $\Ph_1$ which implies
\begin{equation}
\Pt_2\to\Pi_2 \qquad\Longrightarrow\qquad
(1-\Ph_1)\,c_1\,\Ph_1\ =\ 0 \qquad\Longrightarrow\qquad
c_1\,\Th_1\ =\ \Th_1\,Z_1 \quad,
\end{equation}
so that the large-time limit of $\Ph_1$ is also a {\it standard\/} 
one-soliton projector, and we again redenote $\Ph_1\to \Pt_1$. Hence,
\begin{equation}
\lim_{|t|\to\infty} \psi_2\ =\
\Bigl( 1+\frac{\mu_1{-}\bar\mu_1}{\zeta{-}\mu_1}\,\Pt_1 \Bigr)\,
\Bigl( 1+\frac{\mu_2{-}\bar\mu_2}{\zeta{-}\mu_2}\,\Pi_2 \Bigr)\quad,
\end{equation}
which yields
\begin{equation} \label{limphi2}
\lim_{|t|\to\infty} \Phi_2^\dagger\ =\
\Bigl( 1 - \rho_1\,\Pt_1 \Bigr)\,
\Bigl( 1 - \rho_2\,\Pi_2 \Bigr) \quad.
\end{equation}
On the other hand, if we move with the second lump, then, 
in the large-time limit, $c_1\to-(\b_1{-}\b_2)t$, implying
$\Ph_1\to\Pi_1$ where $\Pi_1$ is a coordinate-independent projector.
In this limit, (\ref{Ph1equation}) is satisfied, and (\ref{hatequation})
shows that $\Ph_2{\equiv}\Pt_2$ describes a one-soliton configuration.
Hence, in this frame we find
\begin{equation}
\lim_{|t|\to\infty} \Phi_2^\dagger\ =\
\Bigl( 1 - \rho_1\,\Pi_1 \Bigr)\,
\Bigl( 1 - \rho_2\,\Pt_2 \Bigr) \quad.
\end{equation}

It can be shown, in fact, that the multi-soliton ansatz~(\ref{ansatzpsi})
is equivalent to the product ansatz
\begin{equation}
\psi_m\ =\ \prod_{k=1}^m
\Bigl[ 1 + \frac{\mu_k-\bar{\mu}_k}{\zeta-\mu_k}\;\Ph_k \Bigr] \quad,
\qquad\textrm{with}\qquad
\Ph_k\ =\ \Th_k\,\frac{1}{\Th_k^\dagger\Th_k}\,\Th_k^\dagger \quad,
\end{equation}
where in general the $\Th_k$ are `non-holomorphic' matrices.
By induction of the above argument one easily arrives at the $m$-soliton 
generalization of~(\ref{limphi2}).
Namely, in the frame moving with the $\ell$th lump we have
\begin{equation}
\lim_{|t|\to\infty} \Phi_m^\dagger\ =\
(1-\rho_1\Pi_1)\ldots(1-\rho_{\ell-1}\Pi_{\ell-1})\,
(1-\rho_\ell\Pt_\ell)\,(1-\rho_{\ell+1}\Pi_{\ell+1})\ldots(1-\rho_m\Pi_m) \quad.
\end{equation}
This provides an alternative proof of large-time factorization.

\end{document}